# Quantitative transcriptional analysis of aging *C. elegans*

**Diana David-Rus***

*Department of Bioinformatics and Structural Biochemistry,
Institute of Biochemistry of Romanian Academy,Splaiul independentei 296, 060031,
Bucharest 17, Romania

**Abstract**

My analysis uses methods developed for data mining microarray experiments, adapted for aging research. Methods bridge knowledge of statistical mechanics with data mining methods developed in statistical mathematics. Analyses can reveal how the transcriptional regulation of genes might coincide, thereby implicating proteins as parts of networks acting together towards a common biological function. Such experiments are most useful for complex biological traits that result from the presumed functioning of several molecular pathways. Aging is one such biological phenomenon that apparently incorporates numerous molecular mechanisms underlying environmental stimulus sensing, metabolic regulation, stress responses, reproductive signaling, hibernation, and transcriptional regulation.

## 1. Introduction:

Given the existence of several mechanisms of aging (those conserved across species), simple models have become important tools for elaborating the basic biology of aging. Our lab uses the nematode *C. elegans* to dissect conserved mechanisms of aging.

*Caenorhabditis elegans* was chosen by Sydney Brenner in 1965, as a model organism to study animal development and behavior. This soil nematode has proven to have a great potential for genetic analysis, partly because of its rapid (3-day) life cycle, small size (1.5-mm-long adult), and ease of laboratory cultivation. *C. elegans* natural way of breeding is as a self-fertilizing hermaphrodite. (see Fig. 1- from *C.elegans* II)

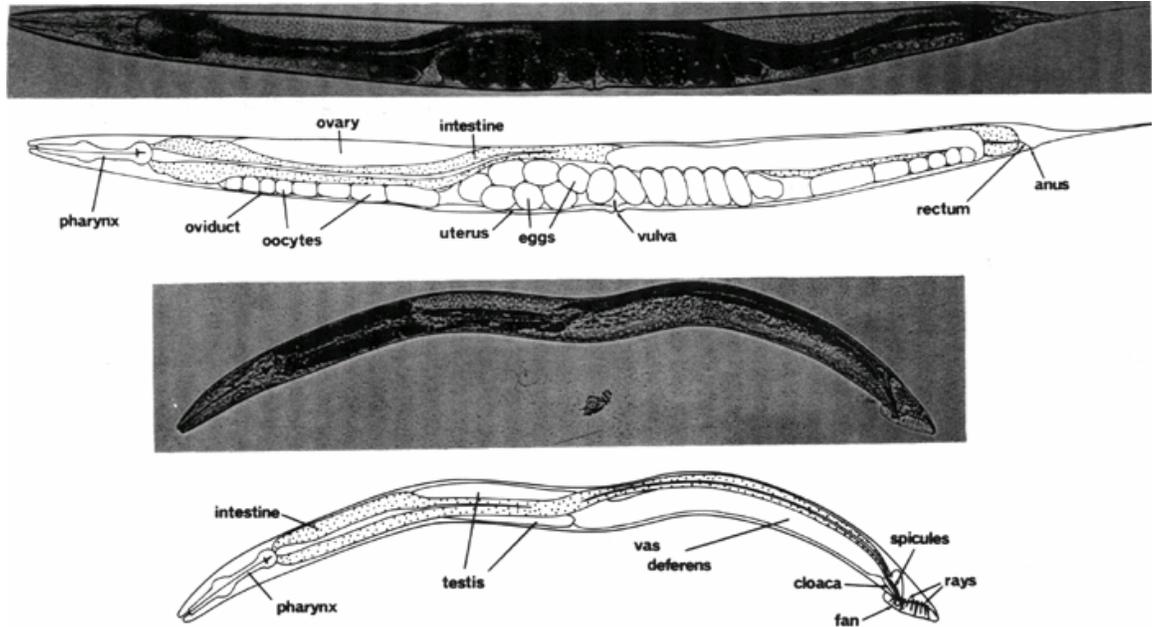

Fig. 1 *C. elegans* anatomy (from from *C. elegans* II book)

Another advantageous feature of this nematode is its size, just 20 times that of *E. coli* and its simple anatomy of 959 cells, including the 302-cell nervous system. With such a small nervous system, *C. elegans* was the first animal model in which its circuitry was completely reconstructed by serial-section electron microscopy (White et al. 1986, 1988). The wild-type reconstructions showed all the connections of all the neurons in the hermaphrodite nervous system. Other unique advantages offered by this organism are the transparency of the body, the constancy of cell number, and the constancy of cell position from individual to individual. Due to such advantages, the complete wild-type cell lineage from fertilized egg to adult was determined by observation of cell divisions and cell migrations in living animals (Sulston et al. 1988).

An essentially complete *C. elegans* sequence was published in *Science* (Vol. 282 11 December 1998) and the last remaining gap in the sequence was finished in October 2002. The completed *C. elegans* genome sequence is represented by over 3,000 individual clone sequences which can be accessed through WormBase. The worm informatics group at the Sanger Institute play an important role in assembling the whole database. WormBase is the repository of mapping, sequencing and phenotypic information for *C. elegans* (and some other nematodes). WormBase is based on the Acedb database system. Acedb was originally developed by Jean Thierry-Mieg (CNRS, Montpellier) and Richard Durbin (Sanger Centre) for the C. *elegans* genome project, from which its name was derived (A C. *elegans* DataBase). However, the tools in it have been generalized to be much more flexible and the same software is now used for many different genomic databases from bacteria to fungi to plants to man. It is also increasingly used for databases with non-biological content.

The entire genome of *C. elegans* encodes approximate 19000 genes. Note that ~50% of *C. elegans* genes have clear human homologs and many molecular processes are strikingly conserved from nematodes to higher organisms. Over 400 mutations or RNAi interventions that extend the ~ 3 week *C. elegans* lifespan have been identified. Although the longevity phenotype is a focal point for much of the work in the field, much less is understood about whether longevity genes actually act to extend *healthspan* (the period of mid-life "vigor" that precedes decline), enabling the animal to live a high quality or "youthful-like" life for longer.

One simple question is what happens to gene expression as animals age--can analysis of transcriptional profiles inform about the biology of aging and suggest ways we might extend healthspan? In the first part of my thesis, I have therefore been analyzing transcriptional profiles of aging animals with a focus on two areas suggested by ongoing work in the Driscoll lab: one addresses the question of whether there is a major 'crisis' or transcriptional transition during midlife, and a second focuses on age-associated muscle deterioration.

## 2  Transciptional profiling to characterize aging and identify genes

## that might impact healthspan

Microarrays are chips on which genes are attached for hybridization to mRNA samples--hybridization signals indicate which genes are expressed as messages and can speak to relative abundance and changes in gene expression over time. We prepared replicate RNA samples over the course of *C. elegans* adult life and hybridized to near complete genome arrays to ask how transcription changes during adulthood and to correlate some of these with aging phenotypes. My analysis uses methods developed for data mining microarray experiments, adapted for aging research. The method I'm using bridges knowledge of statistical mechanics with data mining methods developed in statistical mathematics. Such analyses can reveal how the transcriptional regulation of genes might coincide, thereby implicating proteins as parts of networks acting together towards a common biological function. Such experiments are most useful for complex biological traits that result from the presumed functioning of several molecular pathways. Aging is one such biological phenomenon that incorporates numerous molecular mechanisms underlying environmental stimulus sensing, metabolic regulation, stress responses, reproductive signaling, and transcriptional regulation. Current models of aging emphasize different mechanisms as driving forces behind aging and lifespan determination. However, an integrated understanding of exactly how these mechanisms drive aging has not yet been formulated.

I used supervised and unsupervised methods for gaining a better understanding of the gene expression changes that might impact the aging process. When interpreting the data using a supervised approach, I've tried to address the major biological theories currently known that describe aging. To address the oxidative damage theory of aging, for instance, I highlight stress response genes that exhibit statistically significant changes, and then ask whether the expression patterns of these genes share a common pattern. Overall, my work includes surveys of insulins, longevity-implicated genes, dauer-related genes, autophagy-related genes, muscle, neuronal and germline genes as groups of interest relative to aging and healthspan that I analyze in detail.

Using an unsupervised approach based on concepts from statistical mechanics, I identified an interesting gene expression pattern that suggests that a gene expression switch at midlife. This switch coincides with the onset of biomarkers of aging including age pigment accumulation. I conclude my work with the description of the gene expression sets that underlie new hypotheses about impact on later aging.

Several experiments using microarray technologies to address the different models of aging have been published. A second phase of my analysis was to look at how my data intersects with similar studies performed. I will focus on comparison of just two other experiments due to their similarity with our experiment. I've searched for the overlap between experiments and the common trends in expression pattern using different statistics methods for normalizing and filtering the data.

A first foray into *C. elegans* microarray analysis was performed using Affymetrix oligonucleotide-spotted chips.(Hill et.al. 2000). This study compared gene expression profiles from 18,791 predicted open reading frames, mainly over developmental time points. One mid-life time point (post-fertilization day 14) is included in the study for comparison. One-way ANOVA analysis was performed on the data, which then was normalized to have a mean value of zero and a variance of one. 4221 ORF's with statistically significant variations in frequency (p< 0.001) were identified. Of these ORF's, subclusters (clustered by self-organizing maps (SOM)) of expression patterns that exhibit *declining* expression at the 2-week time point further were studied (Hill et al., 2000).

The first focused microarray study devoted to studying aging in *C. elegans* utilized a probe DNA-spotted microarray of 17, 871 open reading frames to study aging nematodes over a series of time points spanning pre-reproductive adulthood to old age (Lund, et al., 2002). In this study a combination of mutation and strains has been used, see table 1 below:

| Table 1.  Time Course of *C. elegans* from 3 to 19 Days of Age | | | | | |
|---|---|---|---|---|---|
| Day | 3 | 4 | 6–7 | 9–11 | 12–14 | 16–19 |
| *fer-15* | 2 | 0 | 1 | 1 | 1 | 0 |
| *spe-9;fer-15* | 3 | 3 | 2 | 5 | 2 | 3 |
| *spe-9;emb-27* | 1 | 0 | 1 | 0 | 1 | |
| Total = 26 | 6 | 3 | 4 | 6 | 4 | |
| | Young adult | | Oocyte production ends | | | <25% survival |

The number of arrays and the strain of worm included in each time point are shown. Notable characteristics of the population are indicated.

**Table 1  from Lund et. al. 2002**

An ANOVA analysis was undertaken to identify statistically significant variations among the time points, then the data was normalized to the earliest time point (non-aging-related gene expression).  Open reading frames showing variations in expression over time were clustered together in groups showing common expression changes.  Those genes that changed only from the pre-reproductive to first reproductive time point were labeled as maturity genes.  Those genes that had any changes in expression over the successive time points were designated as aging genes.  After statistical filtering, 201 genes exhibited changes over time; 34 maturity genes and 167 aging genes.  Three genes were subsequently discarded due to strong correlation with a particular strain, and 72 of the remaining genes were found to encode proteins conserved across species.

A second study (McElwee et al., 2003) compared the gene expression from *daf-16(-/-)* and *daf-16(+/+)* worms on a *daf-2 (-/-)* reduction of function mutant background, only on the first day of adulthood.  This microarray utilized DNA probes corresponding to 17,871 *C. elegans* genes.  1646 genes were isolated that showed differential expression of greater than 1.5-fold.  602 genes were up-regulated in *daf-16 (+/+),* animals, while 1044 genes were down regulated.

The third study (Murphy et al., 2003) of gene expression changes in aging *C. elegans* utilized DNA probes corresponding to 18455 open reading frames.  This study also compared samples from *daf-2*-deficient and *daf-2;daf-16*-deficient animals, as well as from wild-type animals.  However, this study went further than the McElwee study by comparing the results across multiple time points. The time points begin at a pre-reproductive age and continue until mid-adulthood (later than the time point collected in the McElwee study, but earlier than the Lund et al. study.

A fourth study compares results from microarray studies of aging across species, including *C. elegans* (from the Murphy et al. data), *D. melanogaster*, *S. cerevisiae*, and *H. sapiens*. (McCarroll et al., 2004). In this experiment were performed specific comparisons between *C. elegans* and *D. melanogaster* at two points early adulthood and mid-life adulthood) included several hundred ortholog gene pairs that are conserved in expression across the two species.

A final study is a time-course study of an aging wild-type (N2) and non-aging *daf-2* (-/-). What distinguishes this study is that target samples were prepared from individual nematodes rather than populations, thereby bypassing the variation in aging inherent to worm populations (Golden and Melov, 2004).

Beyond *C. elegans*, other studies have looked at genome-wide transcriptional profiles of aging in specific tissues of other organisms, as well as in the whole organisms of flies and yeast. Such studies have surveyed aging in mouse liver, mouse heart, mouse brain, mouse muscle, rat hippocampus, rat kidney and pituitary, rat muscle, human blood, and human muscle.

When interpreting their data, several of the studies took a similar, supervised approach in the context of current theories of aging. To address the oxidative damage theory of aging, for instance, the studies identify stress response genes that exhibit statistically significant changes, then ask whether the expression patterns of these genes share a common pattern. Conversely, several of the studies have taken an unsupervised approach and examine expression cluster groups for patterns that indicate possible relevance for the biology of aging, looking for commonalities of function among the listed genes, in addition to the relevance of individual genes for aging.

Hill et al. study focuses on cluster groups where the expression patterns decrease with age. One such cluster group was found to be enriched for genes for metabolic activity, including oxidoreductases, amino acid metabolism genes, carbohydrate metabolism genes, and protein synthesis genes. Other down-regulated genes were found to belong to common functional groups, such as muscle-related genes and genes coding for extra-cellular matrix proteins.

In the Lund et al. study, a more supervised approach was taken. Insulins, aging-related genes, dauer-related genes, heat shock genes, transposons, muscle, neuronal and germline genes all were singled out and their expression profiles was examined. Key findings here include: 1) that both aging and dauer-related genes cluster to mount 15 in the Kim expression map when multiple gene expression experiments are combined, 2) while specific insulin genes change in expression over time, the insulin signaling pathway genes do not change over time, 3) both muscle and neuronal genes show an increase in expression in later life, indicative of an up-regulation of the expression of these genes, or indicative of a general down-regulation of the majority of other genes 4) heat shock genes are up-regulated initially, then down-regulated at the latest time points, suggestive of the lack of a stress-response as being possibly causative for the ultimate demise of the organism. Lund results are comparable with our results where we identified heat shock genes as up-regulated as well in cluster G11. By difference with Lund, most of our heat shock genes stay up-regulated over the entire life of the nematode (see more at cluster analysis-G11 cluster) 5) mitochondrial genes and genes involved in oxidative stress resistance do not change over time, 6) transcription of transposable elements is increased, perhaps indicative of a less stable genome with age, and 7) germline genes only are down-regulated at the latest time-point, indicative that the cessation of oocyte production is not dependent on gene expression regulation. Also indicated is that persisting germline tissues into old age may retain functional abilities for reproduction if exposed to a favorable environment. Additionally, an unsupervised approach was taken and 167 genes that show any type of significant change over time are identified. This relatively low number of aging-related genes supports a model where environmental damage contributes more to aging than genetic influences. However, another plausible argument is that the modulation of the expression of a few genes may directly influence life span determination.

In the McElwee et al. study, (Aging cell 2003) results look at comparing an aging (*daf-2 -/-; daf-16 -/-*) and delayed-aging (*daf-2 -/-; daf-16 +/+*) population of worms. In a delayed-aging population, heat shock and oxidative stress-response genes are observed to increase. This increase correlates with the decrease in the heat stress-response genes in the aging Lund et al. population. Furthermore, in the delayed-aging population, metabolic genes are observed to decrease. This finding might speak to the theory that a higher metabolism leads to greater tissue damage and enhanced aging due to buildup of damaging metabolic byproducts. Finally, again corroborating the Lund et al. data, *ins-7* is observed to increase in the delayed-aging population (Lund et al.'s aging population shows a decrease in *ins-7*). Interestingly, no gene expression changes were observed in protein synthesis or protein degradation genes (both proteosomal, and more specific, non-proteosomal genes). This finding does not support the theory that reduced protein turnover in a cell might lead to a buildup of protein damage and an enhancement of aging phenotypes. The Murphy et al. study, is a similar study comparing *daf-16* (+/+)(delayed-aging) and *daf-16*(-/-)(aging), includes many results comparable to McElwee et al.. For example, both oxidative and heat shock stress response genes increase in expression in the delayed-aging population. Furthermore, metabolic genes were decreased in the delayed-aging population. Conversely, however, this study finds *ins-7* decreasing in the delayed-aging (*daf-16 +/+*) population, while increasing in the aging (*daf-16 -/-*) population. I will present later how this study compares with our study. The authors point out further that the gene identified in the McElwee study as *ins-7* really is *ins-30* based on the cosmid name. *ins-18* is up-regulated in the delayed-aging population, in contrast to the up-regulation of *ins-18* in the aging population in the Lund et al. study. Further findings include: 1) that anti-microbial genes are up-regulated in the delayed-aging population. This finding supports a theory that the ultimate demise of the animal is due to bacterial infections overcoming the weakened organism in old age, 2) vitellogenin is down-regulated in the delayed-aging population, consistent with the theory that excessive expression of non-essential genes also may contribute to the aging and demise of the organism, 3) in contrast to the McElwee study, several proteases were repressed in the delayed-aging population, 4) lysosomal genes were up-regulated in the delayed-aging population, and finally 5) genes from the glyoxylate cycle, which are up-regulated during dauer and hibernation, also are up-regulated in the delayed-aging population.  This finding is consistent with the data from Lund et al. that shows dauer and aging genes co-segregating on the same gene expression mountain (Mount 15).

The cross-species comparison between *C. elegans* and *D. melanogaster* (Steven A McCarroll et. al, Nature genetics 2004) reveals trends common to both species or unique to each species.  Trends in common include: a downregulation of many mitochondrial and oxidative metabolism genes (including mitochondrial membrane genes, genes for components of the electron transport chain, ATP synthase genes, and genes in the citric acid cycle), a downregulation of peptidases, proteins for DNA repair, and genes coding for ATP-dependent transporters.  Gene expression changes unique to aging *C. elegans* include:  a downregulation of collagens, histones, trasnposases, and DNA helicases.  Gene expression changes unique to *D. melanogaster* include upregulations of cytochrome p450 genes, glycosylase genes, and peptidoglycan receptors.

A more recent (Golden and Melov, 2004) *C. elegans*  microarray study utilized individual nematodes compared wild- type (N2) to *daf-2*(-/-)  nematodes.  Interestingly, greater gene expression changes were observed between the two strains rather than between different ages of a single strain.  This is consistent with the Lund et al. data that found few changes in gene expression with age.  Many individual genes that relate to current models of aging were identified as different among N2 and daf-2(-/-).  In N2 nematodes, an increase in antimicrobial peptides, mitochondrial electron transport chain proteins, proteasomal components, and actin all can be explained by the earlier onset of aging in N2 worms, thereby increasing the needs for defenses and anabolism to compensate for the age-related deterioration.

While these studies all focus on aging, the varying time points present a problem for the specific study of the mid-life aging, or healthspan, of the organism.  Therefore, we chose to perform a time-course study of aging wild-type nematodes, from reproductive to old aged, with an emphasis on covering the mid-life time points, represented by the post-embryonic days 9-12 when grown at 25 degrees Celsius. Previous studies from our lab reveal that the mid-life changes in an animal may be critical in determining the ultimate lifespan of that animal. All of our samples utilize the same sterile mutant strain and replicates were harvested at the same time point (give or take an hour). Affymetrix oligonucleotide arrays were chosen based upon the good coverage of open reading frames on the array, and based on the optimization of the Affymetrix system. We chose to use a clustering system based upon the statistical mechanics of disordered granular ferromagnets and developed in the Domany lab (M. Blatt, S. Wiseman, and E. Domany, (1996)). This clustering system has proven superior to other clustering methods for a variety of biological problems (E, Domany et.al, (1997, 1998).

## 2.1 A search for mid-life gene expression changes that might influence healthspan- Experimental design

Previously we showed that neuronal cells do not physically deteriorate whereas muscle cells deteriorate morphologically with age, starting in mid-life (see Herndon L. et all Nature 2002).Interestingly my microarray analysis capture such tendencies. I found that neuronal gene expression shows little change for a certain group of neuronal related genes and I showed an overall decline in transcription of muscle expressed genes. Of the muscle related genes, a group of muscle genes is expressed at lower levels during midlife.

Further more, using unsupervised approaches I've identified an interesting gene expression pattern that suggests that a previous unknown genetic switch might occur during midlife day 10 from the time of hatching, consistent with patterning of changes in age pigment accumulation rates. I noted a similar gene expression pattern for day 11 in Kenyon data when I clustered this data (see results at comparison section)

I've also performed a comparative analysis with data from other microarray studies. Here I've looked for the overlap between experiments and the common trend in pattern expression, deploying different statistics methods for normalizing and filtering the data. I've also clustered using Domany algorithm each of the data set I used for comparison.

With an interest in tracking gene expression changes over the *C. elegans* adult lifespan and in identifying genes that are similarly regulated in aging microarray experiments in independent laboratories, we performed microarray analysis using RNA isolated from adults of increasing age.

We identified genes for which expression changes over adult life, using the oligonucleotide type of chip. These chips were specially designed for *C. elegans* by the Hoffmann-LaRoche company, from Basel, Switzerland, as Affymetrics format which effectively covered 87% of the actual predicted genes. The raw data identified by Affymetrix ID annotations is found in the supplementary materials (see Table X).

To grow the worms in a synchronous way, and at the same time limit use of multiple mutations, we used *spe-9*(hc88) which is a temperature sensitive sterile mutation. We cultured *spe-9* mutants at restrictive temperature of 25.5°C. Independent samples containing ~ 20000 worms were taken on different days: day3, day6, day9, day10, day 11, day12, day15. Day 3 is the first day of adulthood, with day 0 the day of hatching. For each day, the mRNA of 3 independent samples was extracted. Each sample was labeled and hybridized to the *C. elegans* genome chip so for each day, we have 3 samples hybridized to 3 independent chips.

Because of our focus was on potential relevant changes at the midlife transition, suggested by changes in rate of age pigment accumulation (Gertsbrein et. al.) we also prepared another triplicate experiment in which we harvested nematodes at days 9, 10 and 11. Data from these middle time-points were combined with those in the more extended trials to increase the significance of findings at days 9, 10 and 11 (we therefore analyzed six total independent repeats for the middle life time points).

## 2.2 **Identifying the 2000 genes that show greatest variance over time points.**

The next steps, called data preprocessing, are important to address several issues related to removal of the effects of systematic sources of variation; to identify if there are still any discrepant observations and to transform the data into a scale suitable for analysis. Preprocessing can greatly enhance the quality of any analysis, therefore is critical to choose the right methods appropriate to the particular type of data and the questions that will be analyzed.

The microarray data can be represented in a matrix form. The rows are the genes covered on our chips, and the 7 columns are the seven time points in which we were interested.

In order to detect the outliers we used the Nalimov outlier test, an outlier exclusion test. For each gene per *condition* a modified Nalimov outlier test (Kaiser R, Gottschalk G (1972)) is performed for data points representing replicate experiments. In contrast to the original test, we used a modified version called "Nalimov1". A normal distribution model is calculated for data points to be tested, and outliers are removed at a 95% confidence level. This means that only in 5 of 100 cases a data point is removed erroneously. Since the test is rather conservative, Nalimov outlier removal normally improves the quality of results, since (chip) artifacts are quite reliably removed. Note that the test requires at least 3 data points (*replicate* experiments) for an experimental condition, otherwise no outliers can be detected. We used, the standard Nalimov 95% confidence level.

Once we identified and discarded the outliers, we scaled the data on each chip and between chips. In order to achieve this, for each chip, we calculated the median signal intensity over all probe sets. The median of this median signal intensity from all chips was calculated. Than, every chip, was scaled to this median value.

The next step was to transform the data using a logarithmic transformation in base 2. The reason we do this is that is preferable to work with logged intensities rather than absolute intensities since the variation of logged intensities tends to be less dependent on the magnitude of the values; taking log reduces the skewness of the distributions, comparing with a Gaussian distribution, and improves variance estimation. Sometimes, "thresholding" is used as part of the preprocessing- any data that have an expression level below the chosen threshold is discarded. We were interested in all the data, since we consider that a low level of expression at a certain time point can be significant for what we were looking for. We therefore didn't consider using any threshold level on our data. We've 'estimated' the data based on the values of the K nearest neighbor genes estimator (see Troyanskaya, T., Tibshirani, R., Botstein, D. & Altman, R. B. (2001)). In this sense I've chosen KNN=12, meaning the range of neighbors for the estimation process is 12. I average these 12 values and replace the smallest value I would like to 'estimate' with that average. This method is considered a better estimator than discarding data through the 'threshold' method.

We also filtered the data. We consider that genes that vary in expression the most over the adult life identify the most regulated genes and therefore tell us more about aging expression of this organisms. Genes were filtered on the basis of their variation across the time samples. We choose a set of 2000 genes that exhibit the greatest variation. In presenting the results of our unsupervised as well as supervised method, we will refer to the list of 2000 genes determined based on highest variation filtering.

The next step for preprocessing the data was normalization. For normalize the data, we performed two steps: first, we centered to the median. We subtracted from each component of the initial vector the median value between the components of that vector, to obtain a new vector. We than normalized the newly obtained vector by dividing each component of the newly obtained vector to its norm i.e.square root of the sum of the squares of the components. By normalizing all genes (or raw in the matrix) we get to the stage of being able to compare the genes with 7 samples between each other and as consequence apply a classification method. Given that we don't know exactly what we expect to find in the data, an unsupervised method is the right method to choose. We choose to cluster the data.

Before clustering the data, we wanted to identify the main direction of variations of the genes, and to get a better understanding of the structure of the data we wanted to cluster. In order to do this, we performed a Principal component analysis (PCA). Using this model one can identify the most important gradient of variation in the data points, identifies the first and second eigenvalues, than rotates the data points such that the maximum variability becomes visible, i.e. by plotting the data on the corresponding first and second eigenvectors.

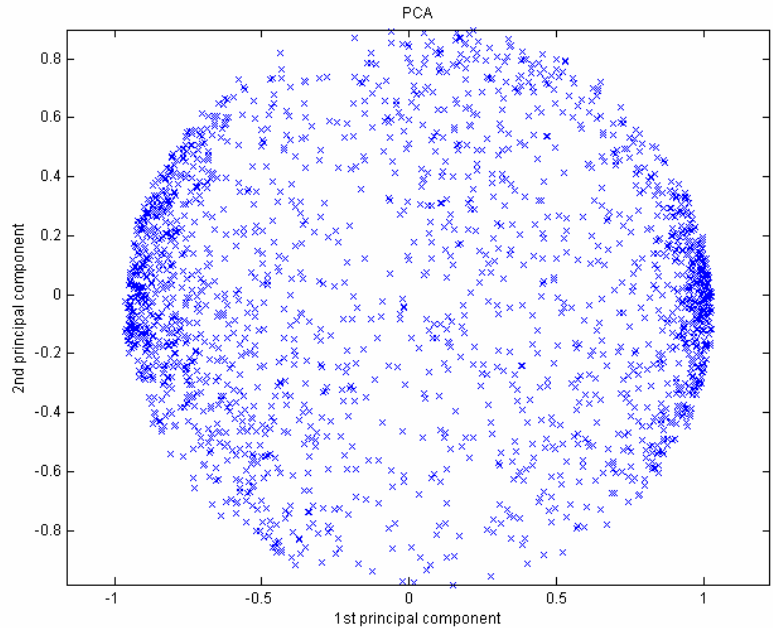

**Fig. 2 Principal component analysis performed on the list of 2000 genes.**

The Principal component analysis provides a first clue of the potential structure existing in the data. For a better understanding of the structure and possible classes existing in the data we will use a clustering algorithm as an unsupervised method. If we would have had already a first classification on the data traditionally we would have used a supervised method. Given that we had no prior classifications on the data, an unsupervised method for classifying the data is required. A supervised method is used when you already have some kind of knowledge on your data, as for example a data classification and you are using that knowledge to learn more from the data, by contrast with unsupervised methods which you are using when you know nothing about the data and you learn first hand from the data. Given that this is our case, we cluster our data.

The clustering method we chose is based on the physical properties of a magnetic system and enables identification of clusters that have not been obtained by other unsupervised clustering methods as Tree- View which are based on Pearson coefficient. This method has a number of unique advantages:

1   Number of the "macroscopic" clusters is an output of the algorithm

2   Hierarchical organization of the data is reflected in the way the clusters split or merge when a control parameter is varied.

3   Being a Monte Carlo based method, the results are insensitive to the initial conditions.

Comparing this SPC algorithm with other clustering algorithms, the draw back of any other methods (such as Tree View) is the high sensitivity to initialization, poor performance when the data contains overlapping clusters; and the most serious problem: lack of cluster validity criteria. None of these methods provide an index that could be used to identify the most significant partitions among those obtained in entire hierarchy. At the same time, we did not want to use a clustering method based on K means algorithm, since this method is known for highly overlapping cluster results that do not necessarily  correspond to  the biological process. The fact that  K means method  can place the  same gene in two different clusters does not necessarily indicate that  one gene can be part of several biological process, but instead reflects incapacity of this algorithm to deal with simply overlapping data (see references for K means). SPC eliminates several of this concerns (see  Domany et.al, (1998)).

Using the SPC algorithm- a Monte Carlo based method, stability, is an attribute of the clusters. The Swendsen-Wang Monte Carlo method has been used, due to its known ability to speed up the algorithm and make it faster. As we look "more deeply" into the data (by increasing a control parameter), and unveil the hierarchical structure of the data, we performed 2500 cycles, with cycle corresponding to a one step increase in control parameter. The number of cycles in which a cluster remains intact, before it is split in other clusters, is called stability. We consider that clusters with higher stability to be more meaningful for biological interpretation of the data (see,   M. Blatt, S. Wiseman and E, Domany, Neural Computation 9, 1805-1842 (1997) for more on algorithm).

## 2.3 Clustering results and interpretation

We clustered the list of 2000 genes using SPC approach to identify 34 clusters, classified, based on size (number of genes in each cluster) and stability.

The hierarchical organization of the data has a graphical representation as a tree, called a dendogram. See fig1 (dendogram_main patterns) with main patterns observed in the data highlighted. The hierarchical organization of the data is reflected in the way clusters split or merge. First, the entire data set of 2000 genes is considered to be part of one giant cluster. As we vary the control parameter the giant cluster will split into multiple small clusters. The clusters or nodes we obtained were annotated as G1-G34, each with a distinctive pattern. The constraints we choose for the clustering algorithm were: minimal cluster 10 (any cluster with less than 10 genes will not be accepted) and stability 3 (meaning, any cluster with stability less than 3 will not be accepted, or, in other words, any cluster which breaks down sooner than 3 cycles).

We performed 2500 cycles on the list of 2000 genes, and we used KNN = 12 (KNN-are nearest neighbors – see M. Blatt, S. Wiseman and E, Domany, Neural Computation (1997), for more on the algorithm, also G. Getz, E. Levine & E. Domany, Department of Physics of Complex Systems, Weizmann Inst., Rehovot, Israel, 2001).

Besides classifications of clusters based on the size and stability criterion mentioned above (found in size/stability table), we attempted a classification based on patterns of gene expression identified in each such cluster. The results of this clustering analysis were compiled for an easy access in a web- based design that facilitates their analysis.

Table 2 below is a sample of the main web page. The rest of the Tables and figures which are web based design can be found in the Supplemental data. Main clusters are displayed each with their respective size, stability and pattern in gene expression. A color coded dendogram based on cluster patterns found is presented in fig1.

Green are all clusters with an up-regulated pattern, yellow, with a down-regulated and pink with a day 10 up-ward peak. Smaller sub-patterns are the orange coded with a down-peak day 10, a grey color for pattern of high peak at day 6 and a lila color senescence pattern,

| Red | high stability; |
|-----|-----------------|
| Yellow | down regulated pattern, cluster break/split from G18 |
| Green | up regulated pattern, cluster break/split from G22 |
| pink | up regulated day 10 pattern |

**Table 2 Color codes corresponding to cluster patterns depicted in the dendogram from Fig. 3**

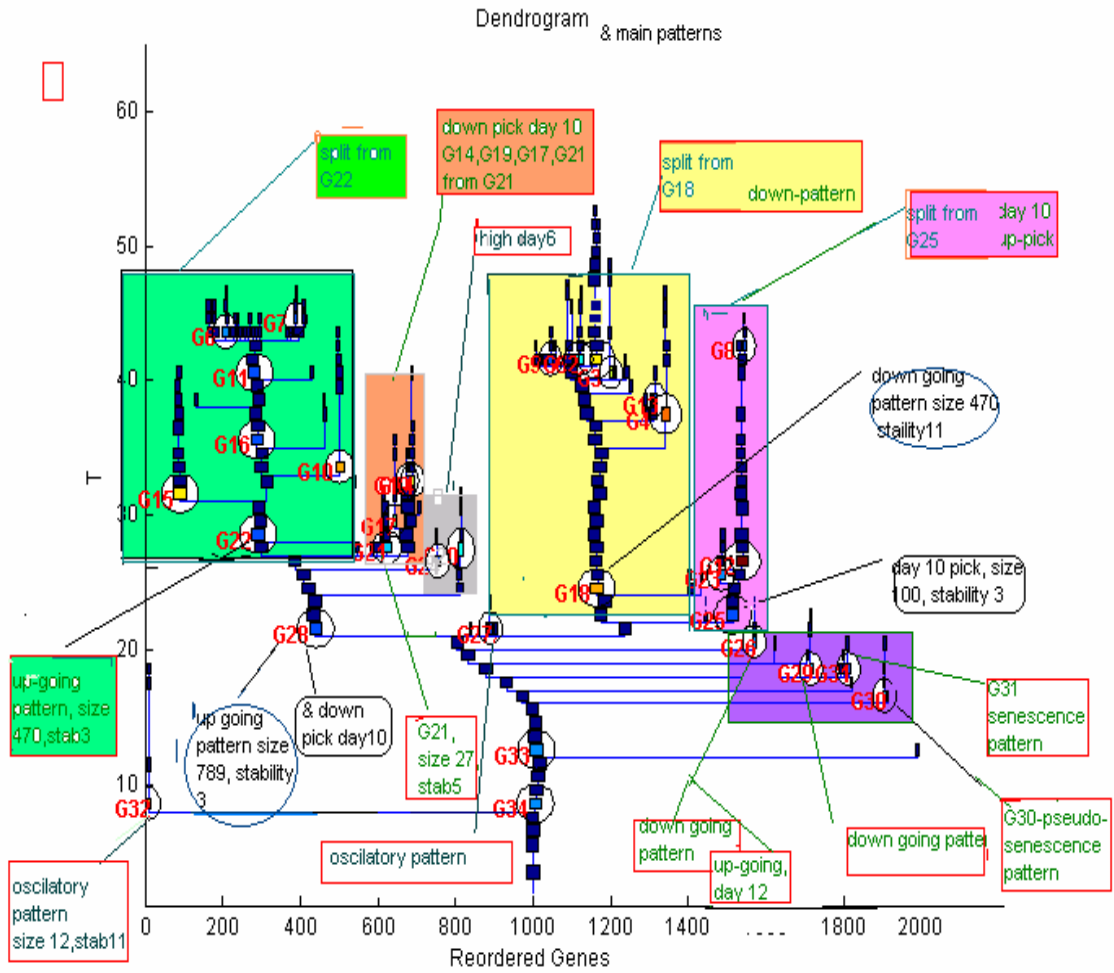

**Fig. 3 dendogram with cluster patterns**

**-TABLE 2-exerpt from the main web page with cluster results from using SPC algorithm**:

| |
|---|
| Clusters of Genes: Clusters are annotated from G1-G34;  G2   Stability=10 Size=63     down regulated pattern (see G18 on the dendogram),   split from G18 |
| G3  Stability=9 Size=13       oscillatory down regulated pattern (see G18 on the dendogram); split from G18;    5 out of 13 collagen |
| G4 Stability=12 Size=24    7 out of 24 collagen; low peak day 6, stay low; split from G18; (see G18 on the dendogram) |
| G5   Stability=6 Size=28       oscillatory down regulated pattern (see G18 on the dendogram); it split from G18,) |
| G6  Stability=4 Size=20    down peak day 10  in upward overall pattern (see G6 on dendogram); it split from G22 |
| G7  Stability=3 Size=11      down regulated peak day 10, in upward overall pattern (see G7 on dendo); it split from G22 |
| G8 Stability=3 Size=11      upward regulated peak day10;             split from G25 |
| G9 Stability=3 Size=12        downward pattern; (split from G18 see dendogram) ) |
| G10 Stability=11 Size=26   high peak day6, left over the rest of *C. elegans* development  split from G22 |
| G11  Stability=3 Size=284      upward regulated pattern;   major size node ;  split from G22 |
| G12 Stability=16 Size=51     upward regulated peak day10;   splits from G25 |

| | |
|---|---|
| G13  Stability=3 Size=11     down regulated peak day10 sub-pattern in down regulated general pattern  (split from G18  see dendogram) | |
| G14 Stability=9 Size=15      down regulated peak day 10       split from G28 | |
| G15 Stability=10 Size=40       up regulated  pattern;      split from G22 | |
| G16 Stability=3 Size=340       up regulated pattern; major size  node     split from G22 | |
| G17  Stability=7 Size=12        down regulated peak day 10      split from G28 | |
| G18  Stability=11 Size=470  down pattern; major size  node from which merge:G2,G3,G5,G9; G4 (dendogram:G18 )      collagen cluster                67 members are collagen related. | |
| G19  Stability=3 Size=11        down peak day 10        split from G28 | |
| G20 Stability=6 Size=12      high- peak day 6, down-going main pattern, down-peak day12 | |
| G21  Stability=5 Size=27     down peak day10;  from G21 splits G14,G17,G19.    G21,splits from G28 | |
| G22  Stability=3 Size=470     up-regulated pattern & low peak day 10; high peak day6 major size node from which merge:                     G6, G7,G10 ,   G11, G15,G16 | |
| G23  Stability=4 Size=27      up-regulated  peak day10;           splits from G25 | |
| G24 Stability=3 Size=14      high day 6, decreasing pattern rest of life | |
| G25  Stability=3 Size=100      up-regulated peak day10;             splits from G25 | |
| G26  Stability=4 Size=11        oscillatory down pattern, senescence pattern-increase day 12-day15 | |

| G27 Stability=3 Size=12          up-peak pattern day 10 |
| G28   Stability=3 Size=789      upward pattern   major size node |
| G29Stability=5 Size=11          down going pattern -senescence as pattern; high expression day 12-day15 |
| G30  Stability=5 Size=10         down peak day 9, pseudo- senescence pattern |
| G31 Stability=3 Size=12         senescence pattern |
| G32  Stability=11 Size=12      oscillatory pattern |
| G33  Stability=4 Size=1941 |
| G34  Stability=4 Size=1978 |

**TABLE 2 extract from main web page, SPC results.**

The entire informational content of the web based clustering design is displayed graphically or in tables. Links from the main web page can be found for:

- Unreordered Data (standardized genes) : a hit-map graph with all the genes normalized before being clustered.

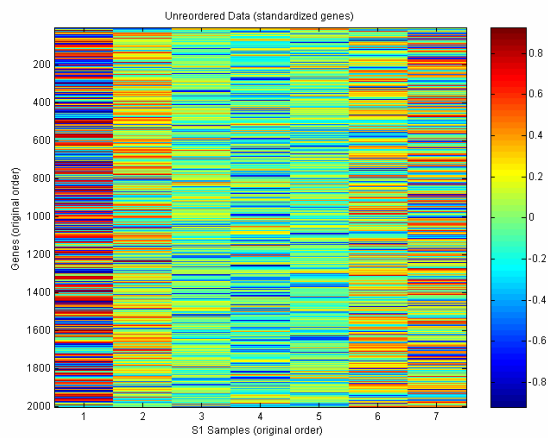

- PCA –a graph displaying the principal component analysis
- KNM –graph with K mutual nearest neighbors; helped in building 'distance matrix'.
- Reordered distance matrix graph-based on which clusters have been identified.

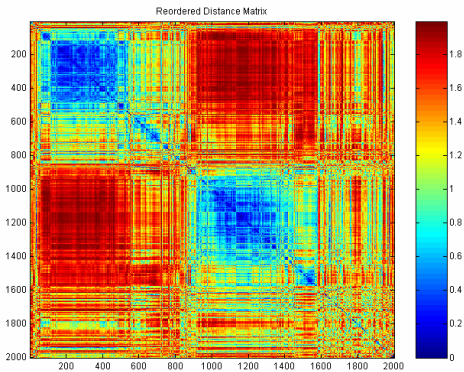

- Dendrogram with Stable Clusters –web based accessible dendogram
- Reordered Data : the entire data list of 2000, reordered after clustering, hit-map graph
- Dendrogram next to Reordered Data : hit-map graph & dendogram

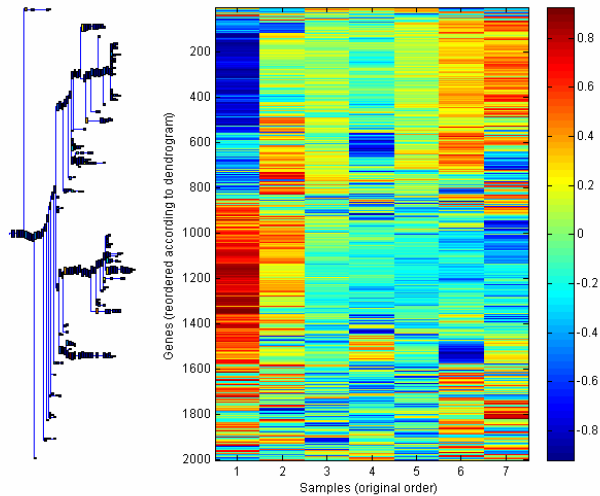

- Reordered Genes : table with all 2000 genes and the clusters where they fit.
- Samples :  time points
- Parameters for SPC

The access to the above is through a link from the main web page (from G1(S1) ). Each cluster can be accessed from the main web page and is represented graphically in two plot formats: as a heat map and as gene expression level changes over time. In addition, a short description of the biological content, of each cluster, correspondence of the cluster with any other clusters, and the list of gene members found in the respective cluster is included. Two tables with clusters sorted based on the stability and size are also presented (see  Tables:  clusters of genes sorted according to stability  and respectively clusters of genes sorted according to size).

 Besides size and stability criterion, the genes in a cluster can be hypothesized to have a functional relationship. An examination of the identity of genes in a cluster can allow the potential nature of this relationship to be addressed. We identified sets of genes that could be grouped by some functional criterion. For example, the genes in some groups shared a protein motif or enzymatic function. In others, the genes were shown to have similar expression patterns or regulation.

 We assume in this experiment we are analyzing the gene expression of the wild type *C. elegans* and that the *spe-9* mutation we used for age synchronicity, has little or no influence on the aging phenotype. The *spe-9* gene is required for fertility in *Caenorhabditis elegans* and encodes a sperm transmembrane protein with an extracellular domain (ECD) that contains 10 epidermal growth factor (EGF) repeats. Evidence suggests that (see Singson A**.** et.al, Dev. Biol.2004) EGF repeats can be mutated to create animals with temperature-sensitive (ts) fertility phenotypes.
.

## 2.3 Clustering results, description and interpretation:

### 1) The general up-regulated pattern: Cluster G28:
**heat-shock genes, insulin like ligands, male- specific genes, genes involved in life expectancy, linkages genes (genes involved in cross-talk).**

 G28 features genes that increase in expression levels in adult life; has the size of 789 genes and stability 3.  This cluster breaks quickly (after 3 cycles) the pattern of up-regulation is maintained in the G22 cluster of size 478 and again, stability 3, and all clusters that merge from it : G15,G16,G11,G6,G7. In the Table 2, as well as in the dendogram  (fig1 above & web based dendogram), these clusters are highlighted in green.

We analyzed the main G28 cluster, the large cluster that groups genes that can be broadly considered to be increasing in expression level as the animals age. An Table with cluster G28 members can be found at Supplemental data. The G28 cluster includes heat shock proteins, insulin-like ligands, and male-specific genes.

Also in this cluster are genes that were found to be expressed at a higher level in long-lived *C. elegans* mutants as compared with  wild-type and short-lived mutants. To determine this, we used the list of short/extended life genes from Murphy et. al. 2003, Nature 424 of ~ 200 genes.  The fact that we see an increase in the expression of the group of genes that are suggested to have a role in the shortening life expectancy of this organism suggests an increase in the relative activity of the products of these genes with obvious consequence of shortening the life of the nematode. On the other hand the increase in the expression of the group of genes involved in long-lived mutants might suggest that beneficial stress resistance genes turn on to protect against aging. Modulating longevity or shortivity genes can impact lifespan

 In the case of heat shock genes, most are found in a single sub-cluster, G11, of size 284 and stability 3 whereas, the other functional groups, are broadly distributed in several different sub-clusters. A Table with cluster G11 members can be found at Supplemental data.

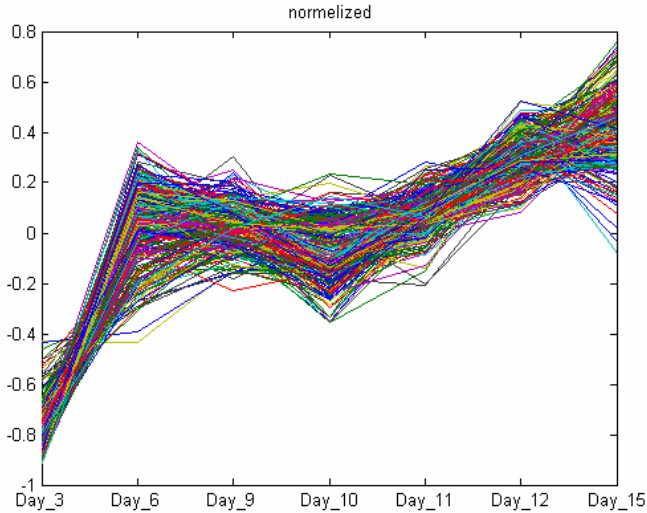

**fig4:G11 cluster mostly heat shock genes**

An increase in the expression of the heat-shock genes might be considered to support the cumulative damage aging theory. The heat shock response gene is a highly conserved biological response, occurring in all organisms. In response to elevated temperature, proteins misfold. As the organism ages, the damaged proteins that misfold accumulates in the organism. Indeed, this might be reflected by the increase in the expression of heat shock genes we are noticing.

**Error-repair mechanism in *C. elegans* doesn't need to be induced with age:**

We should mention here that heat shock genes are not responsible for example for error repair from DNA replication or transcription. Given that we don't notice genes responsible for DNA replication/ transcription repair in the up-regulated cluster G11 we might infer that for the wild type *C.elegans,* this mechanism is not need to be induced in aging animals. We further infer that damage accumulation that might induce aging in the nematode is not due to internal error accumulation like DNA error and transcription accumulation but rather is induced by external stimulus which creates damage that accumulates in organism. The error repair mechanism might work properly and therefore wouldn't require any surplus in gene expression activity. Alternatively, a decrease in expression over time might be required for a proper function of the wild type nematode.

An increase in expression for male-specific genes might be explained by the research done in Andy Singson lab. Interestingly the Singson lab has shown that male mating desire changes with age. One possibility is that the increase in expression of male specific genes might be a signature of age.

**An increase in the activity of the genes involved in "cross-talk" is noticed in the G15 cluster**-this might tell us that with age the communication between cells and in the cells is diminished. Also, from a stability criterion point of view, cluster G15, is interesting since has a high stability of value 10. This cluster has 40 gene members, see table 3 bellow:

| 1 | B0228-5_at | q09433 caenorhabditis elegans. probable thioredoxin. 11/1995 |
|---|---|---|
| 2 | B0284-2_at | best hit: p25386 saccharomyces cerevisiae (baker`s yeast). intracellular protein transport protein uso1. 7/1998 7.0e-10 22% |
| 3 | C08E3-4_at | o17194 c08e3.4 protein. 5/2000 |
| 4 | C17H1-5_at | best hit: o40947 orf 73. 6/2000 2.0e-09 25% |
| 5 | C17H1-6_at | o45257 c17h1.6 protein. 1/1999 |
| 6 | C31B8-4_at | "best hit: q13439 trans-golgi p230 (256 kda golgin) (golgin-245) (72.1 protein) (golgi autoantigen, golgin subfamily a, 4). 5/2000 4.0e-12 25%" |
| 7 | C34D1-3_at | ce08571 locus:odr-3 guanine nucleotide-binding protein (cambridge) tr:q18434 protein_id:cab01489.1. 0/0 |
| 8 | C34E11-3_at | "best hit: p10587 gallus gallus (chicken). myosin heavy chain, gizzard smooth muscle. 12/1998 2.0e-42 22%" |
| 9 | C53A5-9_at | ce08958 ring canal protein like (cambridge) tr:o17700 protein_id:cab03989.1. 0/0 |

| 10 | C53D6-6_at | best hit: q17894 similar to hobo element transposase hfl1. 11/1998 3.0e-27 23% |
|---|---|---|
| 11 | EGAP9-2_at | p91200 cosmid egap9. 5/2000 |
| 12 | F09F9-3_at | q19283 cosmid f09f9. 11/1998 |
| 13 | F11A1-1_at | q19331 f11a1.1 protein. 1/1999 |
| 14 | F13H8-8_at | q19432 cosmid f13h8. 11/1998 |
| 15 | F15A4-9_at | best hit: o01749 similar to human dihydroxyvitamin d3-induced protein. 11/1998 4.0e-17 24% |
| 16 | F26D11-6_at | o61963 f26d11.6 protein. 11/1998 |
| 17 | F26F2-3_at | q9xv56 f26f2.3 protein. 11/1999 |
| 18 | F26F2-4_f_at | q9xv55 f26f2.4 protein. 11/1999 |
| 19 | F26F2-5_i_at | q9xv54 f26f2.5 protein. 11/1999 |
| 20 | F33D11-8_at | o44779 f33d11.8 protein. 11/1998 |
| 21 | F36H5-3_at | p91298 cosmid f36h5. 5/2000 |
| 22 | F43B10-2_at | best hit: p21997 volvox carteri. sulfated surface glycoprotein 185 (ssg 185). 10/1996 3.0e-15 51% |
| 23 | F44G3-8_at | ce16039 f-box domain. (cambridge) tr:o62239 protein_id:cab05520.1. 0/0 |
| 24 | F44G4-6_at | q20416 f44g4.6 protein. 5/2000 |
| 25 | F53B6-4_at | "best hit: p40631 tetrahymena thermophila. micronuclear linker histone polyprotein (mic lh) [contains: linker histone proteins alpha, beta, delta and gamma]. 12/1998 1.0e-14 38%" |
| 26 | F56H6-2_at | o45580 f56h6.2 protein. 5/2000 |

| 27 | F59C6-2_at | best hit: cab92119 dj50o24.4 (novel protein with dhhc zinc finger domain). 7/2000 5.0e-13 37% |
|---|---|---|
| 28 | F59E11-10_at | ce11512 zinc finger protein (st.louis) tr:o16752 protein_id:aab66229.1. 0/0 |
| 29 | H27M09-B_at | "best hit: p40631 tetrahymena thermophila. micronuclear linker histone polyprotein (mic lh) [contains: linker histone proteins alpha, beta, delta and gamma]. 12/1998 4.0e-16 31%" |
| 30 | K07H8-4_at | ce18024 (st.louis) tr:o45179 protein_id:aac04425.1. 0/0 |
| 31 | K09D9-12_at | aaf39930 hypothetical protein k09d9.12. 7/2000 |
| 32 | M162-6_at | ce18896 (cambridge) protein_id:cab05252.1. 0/0 |
| 33 | R07B1-2_at | q09605 caenorhabditis elegans. probable galaptin lec-7. 12/1998 |
| 34 | T07D3-1_at | o16727 t07d3.1 protein. 5/2000 |
| 35 | T22C8-3_at | "ce02351 zinc finger, c2h2 type (cambridge) tr:q22676 protein_id:caa88875.1. 0/0" |
| 36 | W01B6-9_at | best hit: o44929 microtubule binding protein d-clip-190. 6/2000 3.0e-12 22% |
| 37 | Y102A5C-19_at | best hit: o17578 c06h5.2. 5/2000 8.0e-19 29% |
| 38 | Y102A5C-8_g_at | q9xx81 y102a5c.8 protein. 6/2000 |
| 39 | Y53F4A-2_f_at | cab54462 y53f4a.2 protein. 8/2000 |
| 40 | ZK632-11_at | p34656 caenorhabditis elegans. hypothetical 51.8 kda protein zk632.11 in chromosome iii. 11/1997 |

Table 3 G15 cluster members

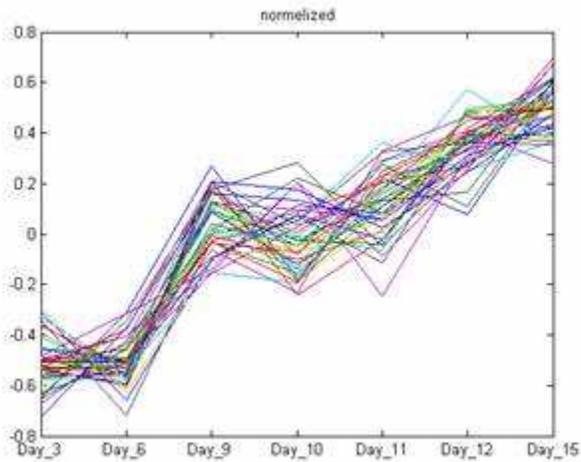

Fig. 5 **G15 cluster** pattern

This G15 cluster has mostly linker and binding type of proteins, intracellular protein transport, canal proteins type, micronuclear linker proteins, linker histones proteins, zink finger proteins, microtubule binder proteins. One reason for this might be that with age the connections and linkages at cellular and intracellular level and the cross-talk in and between cells is weakening therefore an increase in the activity of the products of the genes involved in such processes might be required.

**2) A down-regulated pattern** is noticed in **G18** (**collagen cluster**), **has mostly collagen and muscle related genes.** Size of this cluster G18 is of 470, stability 11. The down regulated pattern is maintained in all clusters that merge from it:G2,G3,G4,G5,G9,G13. In TABLE 2, as well on the dendogram these genes are highlighted on yellow. It is worth mentioning that the G18 cluster has a high stability value 11 in spite of it's size.

The cluster G18 groups genes that can broadly be considered to be decreasing in expression as the animals age. Cuticle collagens are very strongly overrepresented in this cluster—almost one third of the 200+ collagen genes in the worm are found in this cluster.

Kim, et.al (2002 Nature 418: 975-979), identified a large group of genes that are expressed at a relatively high level in muscle tissue; these genes are also in the G18 cluster. Many of the muscle-enriched genes are collagens, but if we exclude collagens from the analysis of muscle genes, we still see most of muscle genes in cluster G18. We also examined 60 genes that are known to have function in muscle (e.g. muscle myosin and other muscle motor proteins), and most of them are also in this G18 cluster. Many of the genes that have human homologs in the total muscle enriched dataset of David Miller(Genome Biology 2007, vol.8, issue 9) are found in this cluster as well. We've used this preliminary results genes for a more careful analysis we've performed later when we've analyzed sarcopenia process in *C. elegans* (see Chapter 2)

**We find two distinct gene expression patterns among genes involved in cellular damage protection: heat-shock genes vs. oxidative stress genes, indicating 2 distinctive gene classes among genes with roles in damage protection.**

We find that peroxidases, cytochrome p450s and glutathione S-transferases are more predominant in cluster G18 with a decrease in gene expression pattern; these types of proteins have a variety of biological functions, but all are commonly involved in detoxification and protection from oxidative stress.

Cellular damage can be induced by heat shock, oxidative stress and various toxic substances. Several pathways are involved in cellular damage. The cumulative theory of aging implies that over time such damage accumulates in organism. Such accumulation might be reflected at the transcription level in an increase in gene expression due to an  increase in the activity required from the genes with a protective role such as heat shock related genes, and observed in the pattern of G11 cluster.

 The finding that genes involved in oxidative stress and detoxification show a decrease in expression pattern opposed to the pattern noticed for heat-shock genes from the cluster G11, might suggest existence of 2 distinct classes among genes involved in cellular damage protection. Misfolding proteins might be associated with an increase in gene expression, as the heat shock  genes in cluster G11 have, whereas the oxidative stress theory of aging might have surprisingly an opposite signature of decreasing in gene expression level as can be depicted in the cluster G18. Possible hypothesis might be that oxidative stress gene activity in the nematode  doesn't have to increase with age given that the oxidative damage from some reason doesn't accumulate with age in *C. elegans*. It would be interestingly to experimentally see the correspondence between, the type of damage these genes are involved i.e type of toxic agent and the pattern and class category enters.

The high stability value of  G18 cluster is maintained in G2,G3 and G4. The G3 cluster contains 5 -collagen related genes out of 13 gene (it's size value), and has an oscillatory down -going pattern. The G4 cluster has an interesting pattern of high expression day3, and low expression rest of the times. It has 24 members genes of which 7 are collagen- related genes (see Fig. 6)

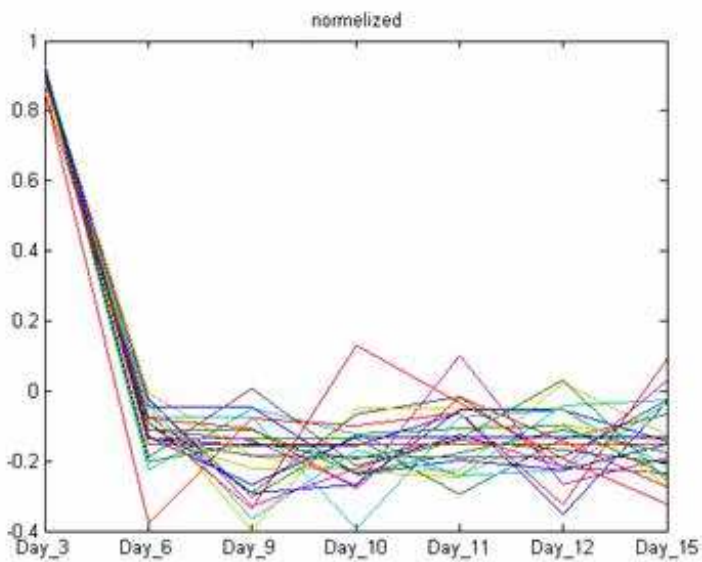

x-axis: time points; y-axis: normalized gene expressions

**Fig.6: G4 cluster pattern**

The G2 cluster has 36 collagen- related genes out of 61 it's total size and has a similar expression pattern as G3.

G2 and G3 clusters show the strongest decline.(see Fig. 7)

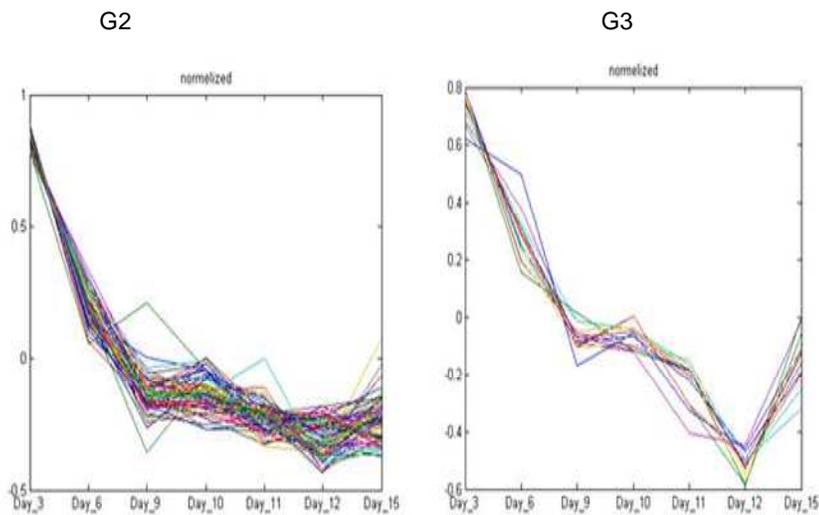

**Fig.7: G2 and G3 cluster patterns; x-axis: time points; y-axis: normalized gene expressions**

**3) The day 10 pattern** can be seen as an oscillatory pattern over all time points with the particularity that for day 10, the pattern of the expression for down or up peak is more pronounced.

**3.a) An oscillatory pattern with an up-peak pattern at day 10 is observed in G25: signaling and transcription factors genes might have a common regulatory loop with germ line genes.**

G25 cluster has size 100, and stability 3, and all the subsequent clusters merge from G25 as: G23, G12, G8. The G25 cluster includes many signaling and transcription factors genes. This group is defined by an inferred role in regulation, e.g. kinases, receptors, G proteins, and the like.

We examined the 23 signaling and transcription factor genes in G25, and found that 7 of them had well-characterized functions in the germline or in early embryonic development. The G23 cluster has the most prominent day 10 change pattern.

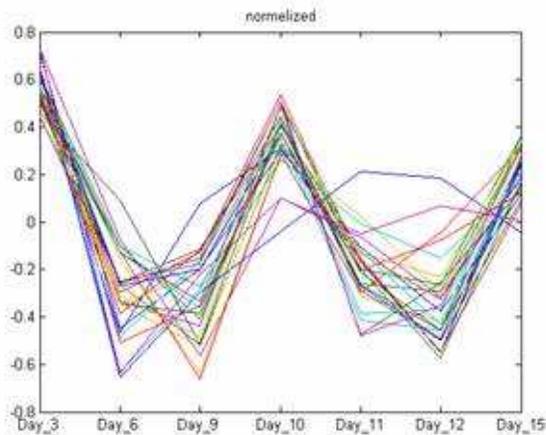

Fig. 8   **G23 cluster oscillatory pattern with a day 10 peak** pattern;  **x-axis: time points; y-axis: normalized gene expressions**

The fact that an oscillatory pattern describes signaling and transcription factors genes is expected. What was a surprise for us was that this oscillatory pattern has high expression peak for day 10. When we later examined a group of germ line enriched genes we find that they have same gene expression pattern. In case that this oscillatory day10 up regulated pattern might indeed be a germ line enriched gene signature than all the rest of genes in this G25 cluster as signaling and transcription factors genes might have a common regulatory loop with the germ line genes. See also the discussion on the germ line genes .

**3.b) an oscillatory pattern with a pronounced down regulation pattern for day 10**.

This pattern can be seen first in G21, and then the sub-clusters G14,G17,G19. This down peak pattern at day 10 can also be seen in the clusters with a general, up-going trend as pattern, as in clusters G6, and G7, which merged from G22 (the main up-going pattern); as well as in the clusters with down-going general pattern as can be seen in the G13 cluster, which merge from G18.

All the clusters with the down peak pattern at day 10 can be seen in Table2 as well as in dendograms fig 3 (highlighted in light pink). G14 is the cluster with highest stability among cluster with down going day 10 pattern (size 15 and stability 9). G14 contains the expression of sri-29 chemoreceptor, gcy-21 which is a protein kinase, fmo-1 which has disulfide oxidoreductase activity and monooxygenase activity, also genes with proteolysis function. See Table 4 bellow for cluster members also fig. 9 for cluster pattern.

| | | |
|---|---|---|
| 1 | B0454-4_at <br><br> sri-29 | o17170 b0454.4 protein. 6/2000 <br><br> chemoreceptor, sri family <br><br> - (*Serpentine Receptor, class I*) |
| 2 | F08E10-2_r_at | q9xxp2 f08e10.2 protein. 5/2000 |
| 3 | F22A3-4_at | ce04440 contains similarity with human homeotic protein pbx2 (st.louis) tr:q19696 protein_id:aaa83195.1. 0/0 |
| 4 | F22E5-3_g_at <br><br> or gcy-21 | ce09555 locus:gcy-21 protein kinase (st.louis) tr:o16715 protein_id:aab66169.1. 0/0 |
| 5 | F36H12-16_at | o76718 f36h12.16 protein. 11/1998;contains similarity to Lactobacillus delbrueckii Abc transporter ATP-binding protein |
| 6 | F53G2-1_at | best hit: q23181 similarity to c.elegans early embryogenesis |

| | | zyg-11 protein. 5/1999 4.0e-66 29% |
|---|---|---|
| 7 | K08C7-2_at<br><br>or fmo-1 | ce21038 dimethylaniline monooxygenase (cambridge) tr:q21311 protein_id:caa94291.1. 0/0;<br><br>dimethylaniline monooxygenase (N-oxide-forming) activity<br><br><br>disulfide oxidoreductase activity; monooxygenase activity |
| 8 | T06A1-5_at | best hit: q21003 similarity to a putative single-stranded nucleic acid binding protein. 11/1998 5.0e-47 29%<br><br>contains similarity to Pfam domain PF01697 (Domain of unknown function)<br><br>molecular_function unknown |
| 9 | T23B12-5_at | similarities with: p70561 fgf receptor activating protein frag1. 8/1998 3.0e-09 27%, is a gene from Ratus norvegicus;<br><br>FRAG1, a gene that activates fibroblast growth factor receptor by C-terminal fusion through chromosomal rearrangement."; |
| 10 | W07B8-1_at | ce14674 thiol protease (st.louis) tr:o16289 protein_id:aab65343.1. 0/0<br><br>proteolysis and peptidolysis |

| 11 | Y47D7A-F_at | aaf60634 hypothetical protein y47d7a.f. 7/2000 |
| 12 | Y48E1B-6_at | o18200 y48e1b.6 protein. 1/1999 |
| 13 | Y54G9A-1_at | q9xwh1 y54g9a.1 protein. 11/1999 |
| 14 | Y54G9A-2_at | q9xwh2 y54g9a.2 protein. 11/1999<br><br>contains similarity to Giardia lamblia Median body protein.; |
| 15 | ZK250-5_at | o17299 zk250.5 protein. 5/2000 |

**Table 4- G14 cluster member. -Some of the genes have direct links to worm base. The yellow highlighted gene has human similarities**

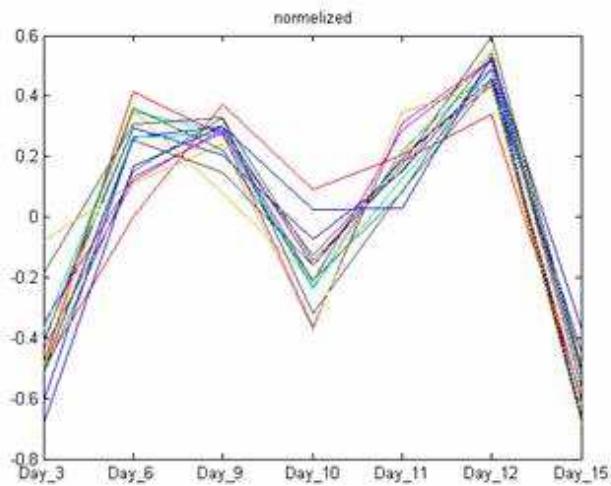

**Fig. 9  pattern of cluster G14:**

**x-axis: time points; y-axis: normalized gene expressions**

As mentioned these genes are involved in signaling or are transcription factors.

**Our data reveal a dramatic change in gene expression around day 10 in at least 13 clusters.**

The consistency of the day10 pattern suggest a significant physiological transition in the *C. elegans* organism during the middle life span time window day 9- day12 of the nematode. Given that some clusters have an oscillatory up-regulated day 10 pattern and others have a down-regulated day 10 pattern might suggest existence of an complementory process among signaling and transcription factors genes involved in the up-regulated day 10 pattern versus the genes involved in the down-regulated pattern.

The reason we were able to narrow down this middle life time window changes focused arround day 10 is due to the statistical analysis performed, as well as the design of the experiment. This day 10 pattern persists even when we repeated the experiment We note that day 10 is measured from day0, the moment of hatching, that the worms were grown at 25C in this study.

**Sub-clusters patterns:**

- **Senescence pattern, neuronal related genes:**

Besides this main patterns defined by large clusters we will mention two other patterns described by smaller sized clusters and found in the list of 2000 adult regulated genes obtained after filtering.

One such pattern is a "senescence" pattern. The cluster with such a pattern is G31. This pattern can be characterized as a relative low, constant level of expression that spans the life of the *C. elegans* from day 3, toward the end of life of the nematode at day 12 with a drastic increase in expression level between day 12 and day 15. This last day is when most of the nematodes grown at the temperature of 25C have already died and the viable animals we assigned are all decrepit. Day 12 is the time when the decay of *C. elegans* as an organism is easily noticeable.

The cluster G31 includes heat shock genes, a homolog of human fetal brain protein, olfactory receptor, stress-inducible protein and sodium neurotransmitter see Table 5 below with cluster G31 members:

| 1 | C01G6-9_at | q17575 c01g6.9 protein. 1/1999 |
|---|---|---|
| 2 | C12D8-4_at | ce05268 transthyretin-like family (cambridge) tr:q17937 protein_id:caa98234.1. 0/0 |
| 3 | C49A9-5_at | o44151 c49a9.5 protein. 11/1998 |
| 4 | EGAP1-1_at | q19073 cosmid egap1. 11/1998 |

| 5 | F08H9-4_at | ce09268 heat shock protein (cambridge) tr:q19228 protein_id:cab01147.1. 0/0 |
|---|---|---|
| 6 | F25H5-7_g_at | ce15903 protein-tyrosine phosphatase (cambridge) tr:o17840 protein_id:cab02988.1. 0/0 |
| 7 | F59B2-9_at | ce00236 f-box domain. (cambridge) sw:p34484 protein_id:caa77586.1. 0/0 |
| 8 | T07D4-2_at | best hit: q15777 homo sapiens (human). fetal brain protein 239 (239fb). 5/2000 4.0e-50 41% |
| 9 | Y37A1C-1B_r_at | q9xtc4 y37a1c.1b protein. 11/1999 |
| 10 | Y61B8A-1_at | best hit: p91118 similarity in c. elegans olfactory receptor odr-10. 6/2000 4.0e-21 31% |
| 11 | ZK1010-9_at | ce23490 sodium:neurotransmitter symporter (cambridge) tr:o18288 protein_id:cab04975.1. 0/0 |
| 12 | ZK328-7_at | ce05072 stress-inducible protein sti1 (st.louis) tr:q23468 protein_id:aaa91253.1. 0/0 |

**Table 5 G31 cluster members**

This senescence pattern (See fig. 10 cluster G31 pattern) was also observed when we made the analysis of the previous data with just 3 replicates per each time point.

Note that G31 cluster has a majority of neuronal genes in the form of neurotransmitters, human homologies as fetal brain or olfactory genes. Also comparing with the sub-pattern of 'young adult-day 6' note that the neurotransmitter involved in the senescence pattern is Na-related by comparison with the day 6 young-adult cluster which has K-related neurotransmitter. See the gene members in G10 cluster.

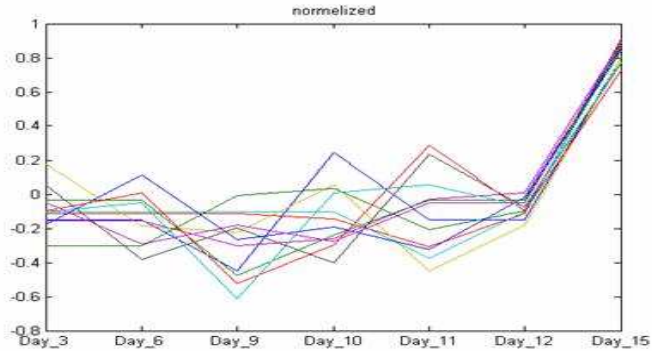

**fig. 10 cluster G31; x-axis: time points; y-axis: normalized gene expressions**

Both "day 10 pattern" as well as the "senescence pattern" are patterns that were never noticed in previous experiments performed by other labs. We consider that this is due in part of the design of our experiment, in particular, the way we chose the time point sampling, and in part, of the statistical analysis performed, in particular the  clustering algorithm used.

- **‘Young adult pattern, - day6 pattern’-has genes specific for larval development and adult morphogenesis and again neuronal  genes.**

    **The K-related channel is member of the young adult cluster by difference with Na-channel related which can be found in ‘senescence pattern’ cluster.**

We found in G10 an opposite pattern from the senescence pattern as day 6 pattern

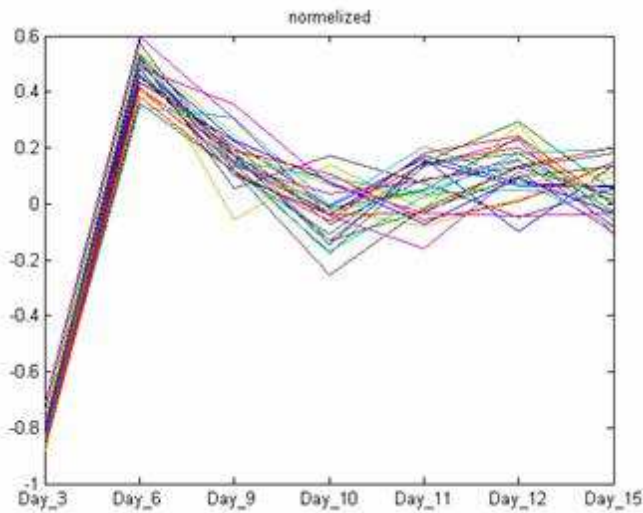

**fig. 11 cluster G10; x-axis: time points; y-axis: normalized gene expressions**

The G10 cluster contains kqt-1 a human homolog, which encodes one of three *C. elegans* KCNQ-like potassium channel subunits that, with respect to humans, is most similar to the KCNQ2-5 subfamily of channel proteins; genes similar to human necrosis factor-alpha-induced protein with voltage gated potassium channel activity. It might be important to note that 'young adult day-6' cluster contains K-channel related genes by difference with 'senescence' cluster which has Na-channel related gene. Also both clusters G31 and G10 have a gene expression similar with protein tyrosine phosphatase. Otherwise, G10 also includes genes specific for larval development and morphogenesis as insulin like families that affect dauer formation and eating behavior like *ptr-3* gene which is in same family with *daf-6*; *acn-1*, required for larval development and adult morphogenesis; the hypodermal expression of *acn-1* appears to be controlled by *nhr-23* and *nhr-25*. Another dauer related gene is *crb-1* which affects dauer formation and eating behavior. Below is Table 6 with members of cluster G10.

| | | |
|---|---|---|
| 1 | C16C8-9_at | p91047 cosmid c16c8. 11/1998 |
| 2 | C18A3-8_at | ce01800 helix-loop-helix transcription factor (st.louis) tr:q09961 protein_id:aaa68375.1. 0/0 |
| 3 | C25B8-1 at or kqt-1 | ce08386 locus:klq-1 voltage-gated potassium channel (st.louis). 0/0 <br><br> The kqt-1 gene encodes one of three C. elegans KCNQ-like potassium channel subunits that, with respect to humans, is most similar to the KCNQ2-5 subfamily of channel proteins; |
| 4 | C33E10-1 at | best hit: q18385 similar to protein tyrosine phosphatase. 6/2000 1.0e-11 25% |
| | | encodes an protein containing an F box (motif considered to mediate protein/protein interaction) |
| 5 | C40A11-3 at | best hit: p91563 similar to human necrosis factor-alpha-induced protein b12. 6/2000 1.0e-34 45%<br><br>voltage-gated potassium channel activity |
| 6 | C41D7-2 at<br><br>or ptr-3 | best hit: p91184 similar to c. elegans protein f44f4.4. 6/2000 3.0e-93 29%<br><br>from PaTched Related family; in this family is ptr-7 as well, known as daf-6 |
| 7 | C41H7-5 at | best hit: q21396 similarity to c. elegans proteins c18h2.1 and t28d9.9. 11/1998 1.0e-20 25%<br><br>also,(as second hit )similarity with gene CBG19613 from C. briggsae which has similarity with SGD:YKL129C from S.cerevisiae which is<br><br>an class I myosin; One of two class-I myosins; localizes to actin cortical patches; deletion of MYO3 has little affect on growth, but myo3 myo5 double deletion causes severe defects in growth and actin cytoskeletion organization; myosin I |
| 8 | C42D8-5 at<br><br>or acn-1 | ce06951 peptidase (st.louis) tr:q18581 protein_id:aaa98719.1. 0/0<br><br>acn-1 encodes an ACE-like protein required for larval development and adult morphogenesis, |
| | | is expressed in hypodermal cells, vulval precursor cells, and ray papillae in the male tail; the hypodermal expression of acn-1 appears to be controlled by nhr-23 and nhr-25.<br><br>==acn-1(RNAi) animals have arrested larval development== |
| 9 | C56E6-6_at | ce04278 leucine-rich repeats (st.louis) tr:q18902 protein_id:aaa81094.1. 0/0<br><br>==similarity with H. sapiens Insulin-like growth factor== binding protein complex acid labile chain precursor<br><br>and with ==S.cerevisiae protein required for  START A of cell cycle== |
| 10 | EEED8-6_at | "best hit: baa91749 cdna flj10682 fis, clone nt2rp3000072. 7/2000 2.0e-57 40%"<br><br> biological fct:proteolysis and peptidolysis<br><br>molecular fct: carboxypeptidase A activity |
| 11 | F02C12-3_at | ce23626 (cambridge) tr:q19109 protein_id:caa91020.2. 0/0 |
| 12 | F11C7-4_at<br><br>or crb-1 | ce07053 egf-like repeats (st.louis) tr:q19350 protein_id:aac69012.1. 0/0<br><br>proteins with same EGF-like domain are: UNC-52 plays essential roles in muscle structure development and regulation of growth factor-like signaling pathways; other genes which encode for proteins with EGF |

| | | like domain:eat-20,spe-9,mec-9 and so on;<br><br>crb-1 encodes a homolog of Drosophila CRUMBS that <mark>affects dauer formation and feeding behavior</mark> |
|---|---|---|
| 13 | F11D11-3_at | o62153 f11d11.3 protein. 1/1999 |
| 14 | F11D11-3_g_at | o62153 f11d11.3 protein. 1/1999 |
| 15 | F23D12-5_at | best hit: o14607 homo sapiens (human). ubiquitously transcribed y chromosome tetratricopeptide repeat protein (ubiquitously transcribed tpr protein on the y chromosome). 7/1999 5.0e-50 28% |
| 16 | F26D10-12_at | ce19812 lectin c-type domain (cambridge) protein_id:cab02321.1. 0/0 |
| 17 | F27E11-1_at | ce09732 nucleoside transporter (st.louis) tr:o16192 protein_id:aab65255.1. 0/0 |
| 18 | F32H2-6_at | ce09881 fatty acid synthase (n-terminus) (cambridge) tr:p91866 protein_id:cab04239.1. 0/0 |
| 19 | F55C9-3_at | q9xuy9 f55c9.3 protein. 11/1999 |
| 20 | F55C9-5_at | q9xuz0 f55c9.5 protein. 5/2000 |
| 21 | H16D19-4_at | q9xx93 h16d19.4 protein. 11/1999 |
| 22 | K08B4-2_at | caeelgn; k08b4-2; -. 7/100; similarity with SW:O35598 from M. Musculus, contributes to the normal cleavage of the cellular prion protein. |
| 23 | R13H4- | p90946 r13h4.7 protein. 1/1999 |

| | 7_at | |
|---|---|---|
| 24 | T02E9-5_at | q9u382 t02e9.5 protein. 5/2000 |
| 25 | T14B4-6_at  dyp-2 or rol-2 | p35799 caenorhabditis elegans. cuticle collagen dpy-2 precursor. 11/1997 |
| 26 | W04A8-4_at | q9xul8 w04a8.4 protein. 11/1999 |

**Table 6: G10 cluster members; the highlighted genes might be representative for the biological theme of the G10 cluster.**

**Higher gene expression after day 6 for the gene members of 'day-6 young adult' cluster pattern might induce shortening in life span of the nematode.**

A high peak at day 6 and than decreasing pattern for the rest of time points, is also maintained in the G24 cluster of size 14 and stability 3. G24 contains *unc-44*, a collagen related gene, and *ces-2*, which is required to activate programmed cell death in the sister cells of the serotoninergic neurosecretory motor (NSM) neurons, and is transcriptionally inhibited by activated LET-60. We can hypothesize that the genes in cluster G10 and G24  are genes important for  adult morphogenesis; cell growth and in general genes involved in cell homeostasis.

Considering the cluster pattern of high peak at day 6 and than a general steady gene expression for the rest of the life of this nematode we might consider that gene members of this clusters might act as 'left-on' genes.

**Day 6- young adult pattern + senescence pattern**

G20 cluster is interesting because it has the high peak pattern at day 6, and the "senescence" pattern, of increase at end stages as well. It contains transcription factors and heat shock protein. See the gene pattern and gene members in  fig. 10 and respectively Table 7 below

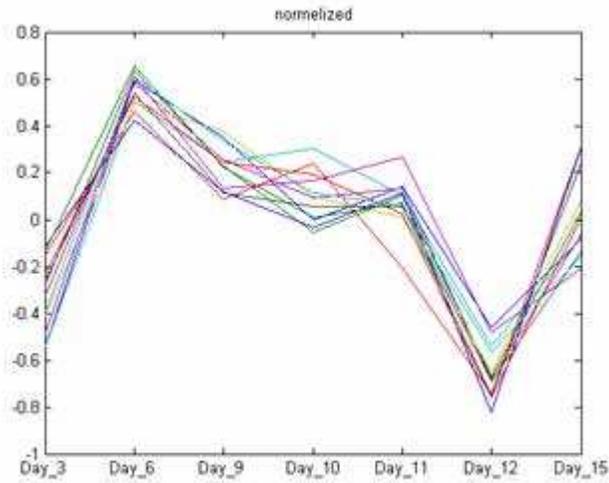

**fig.12  G20 cluster; x-axis: time points; y-axis: normalized gene expressions**

| 1 | C09G12-5_f_at | o44457 c09g12.5 protein. 11/1998 |
|---|---|---|
| 2 | C09G12-5_i_at | o44457 c09g12.5 protein. 11/1998 |
| 3 | C50F4-3_at | ce05468 thiol protease (cambridge) tr:q18740 protein_id:caa94738.1. 0/0 |
| 4 | C54F6-8_at<br><br>**C54F6.8** | ce17603 nuclear hormone receptor (st.louis) tr:o16443 protein_id:aab65942.1. 0/0;<br><br>transcription factor:; associated with:<br><br>• **asp-4--**biological fct :induction of non-apoptotic programmed cell death!<br>• **csn-3** -biological fct(b.f.):control development;<br>• **csp-3** –b.f.:destruction of protein or peptides by hydrolysis (proteolysis, peptidolysis)<br>• **clp-3** –involved in neurodegeneration caused by |
| | | necrotic cell death<br><br>• **unc-71** - proteolysis and peptidolysis; |
| 5 | F02E11-5_at | best hit: o44932 vespid allergen antigen homolog. 6/2000 2.0e-23 35% |
| 6 | F35D11-8_at | q20037 cosmid f35d11. 11/1998 |
| 7 | F35H8-2_at | best hit: q22481 similarity to c. elegans hyuopthetical protein. 11/1998 4.0e-12 32% |
| 8 | F40G9-7_at | q9tz78 f40g9.7 protein. 5/2000 |
| 9 | F59D6-1_at | o16344 f59d6.1 protein. 11/1998 |
| 10 | K03D3-5_at<br><br>**K03D3.5** | • o45643 k03d3.5 protein. 1/1999<br><br>• **K03D3.5-** by blast- best match with a heat shock protein from of B. aphidicola organism |
| 11 | T02D1-7_at | o45726 t02d1.7 protein. 1/1999 |
| 12 | ZK1290-1_at | q23439 cosmid zk1290. 11/1998 |

**Table 7 G20 cluster members**

**G4-Young adult-day3-day6 pattern:**

G4, splits from G18 with a distinct pattern. G4 has a size of 24 expression genes, and stability 12. It starts with a high pattern at day 3, and than a relative constant low expression. G4 has 7 collagen related genes. Because of high expression pattern at day 3 the genes in this cluster might play an important role in the young adult life of this nematode-see Table 8 with gene members of G4 cluster as well as Fig. 11 for cluster pattern.

| 1 | B0024-2_at | ce05147 collagen (cambridge) tr:q17418 protein_id:caa94875.1. 0/0 |
|---|---|---|
| 2 | C09G5-6_at | q09457 caenorhabditis elegans. putative cuticle collagen c09g5.6. 11/1997 |
| 3 | C29E4-1_at | p34340 caenorhabditis elegans. putative cuticle collagen c29e4.1. 11/1997 |
| 4 | E03H12-2_at | o02128 cosmid e03h12. 11/1998 |
| 5 | F11E6-2_at | q9u3j9 f11e6.2 protein. 5/2000 |
| 6 | F12E12-C_at | best hit: q17724 similar to the insect-type alcohol dehydrogenase/ribitol dehydrogenase family. 5/2000 2.0e-73 55% |
| 7 | F36A4-11_at | q20088 similarity to collagen. 11/1998 |
| 8 | F36A4-1_at | q20096 cosmid f36a4. 11/1998 |
| 9 | F37B1-6_at | ce09993 glutathione s-transferase (cambridge) tr:q93699 protein_id:cab02292.1. 0/0 |
| 10 | F40A3-6_at | o16266 f40a3.6 protein. 11/1998 |
| 11 | F41G4-1_at | best hit: q26630 axonemal dynein light chain p33. 11/1998 6.0e-41 44% |
| 12 | F42A10-7_at | q20312 cosmid f42a10. 11/1998 |
| 13 | M18-1_at | ce06193 collagen (cambridge) tr:q21556 protein_id:caa92826.1. 0/0 |
| 14 | R09A8-4_at | ce03540 cuticle collagen (cambridge) tr:q21855 protein_id:caa92006.1. 0/0 |
| 15 | T01B10- | o02153 cosmid t01b10. 11/1998 |

| | 2_at | |
|---|---|---|
| 16 | T10E10-G_i_at | q22326 similar to <mark>collagen</mark>. 6/2000 |
| 17 | T11F9-8_at | ce06420 zinc metalloprotease (cambridge) tr:q22400 protein_id:caa98532.1. 0/0 |
| 18 | T20B3-2_at | ce20087 troponin (cambridge) protein_id:cab04737.1. 0/0 |
| 19 | W09B7-B_at | aaf60391 hypothetical protein w09b7.b. 7/2000 |
| 20 | Y57A10B-6_at | best hit: p41991 caenorhabditis elegans. pes-10 protein. 11/1995 2.0e-38 28% |
| 21 | ZC101-2E_at | q06561 caenorhabditis elegans. basement membrane proteoglycan precursor (perlecan homolog). 7/1999 |
| 22 | ZC373-6_at | q23262 zc373.6 protein. 1/1999 |
| 23 | ZK1193-3_at | q23410 similarity over a short region to tenascin precursors. 5/2000 |
| 24 | ZK1193-4_at | q23412 cosmid zk1193. 11/1998 |

Table 8 G4 cluster members

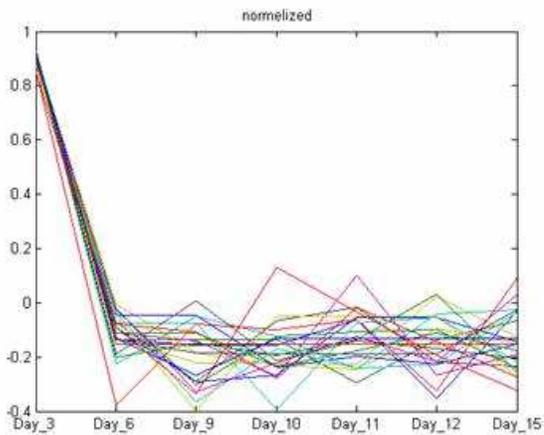

**fig.13 G4 cluster; x-axis: time points; y-axis: normalized gene expressions**

**<u>Summary clusters result:</u>**

We've analyzed and present some clusters that include the 5 major expression patterns. We put more emphasis on clusters with high stability. Any other clusters that are not discussed here can be found in Table clusters in the Appendix for Chapter 1.

In Table 10 is a summary of the clusters with a short description and color based representation of the cluster pattern.

| <span style="color:red">**Red**</span> | <span style="color:red">**high stability;**</span> |
|---|---|
| <mark>**Yellow**</mark> | <mark>**down regulated pattern, cluster break/split from G18**</mark> |
| <span style="color:green">**Green**</span> | <span style="color:green">**up regulated pattern, cluster break/split from G22**</span> |
| <span style="color:magenta">**pink**</span> | <span style="color:magenta">**up regulated day 10 pattern**</span> |

**Table 9 legend for the color based representation in the Table below with cluster summary .**

| |
|---|
| <mark>**<u>G2</u>**</mark> <mark style="background:red">**Stability=10**</mark>**Size=63     down pattern (see <span style="color:blue"><u>G18</u></span>  on the dendogram)** <mark>**split from G18**</mark> |
| <mark>**<u>G3</u>**</mark> <mark style="background:red">**Stability=9**</mark>  **Size=13      oscillatory down  pattern  (see  <span style="color:blue"><u>G18</u></span>  on  the dendogram;** <mark>**split from G18;**</mark> |
| **5 out of 13 collagen** |

**G4** **Stability=12** Size=24    7 out of 24 collagen; **low pick day 6,stay low**; **split from G18; (see G18** on the dendogram)

**G5** Stability=6 Size=28        oscillatory down pattern (see G18 on the dendogram); it **split from G18,**)

**G6** Stability=4 Size=20   **down pick day 10**  in upward overall pattern(see G6 on dendogram); **it split from G22**

**G7** Stability=3 Size=11   **down pick day 10,**in upward overall pattern( see G7 on dendo); **it split from G22**

**G8** Stability=3 Size=11        upward pick day10;        **split's from G25**

**G9** Stability=3  Size=12        downward  pattern;(  it  **split fromG18**see dendogram**) )**

**G10** **Stability=11** Size=26   **high pick day6,** left over the rest  of C.elegans development **split from G22**

**G11** Stability=3 Size=284     **upward pattern;** **major size node ;** **split from G22**

**G12** **Stability=16** Size=51     upward pick day10;     **split's from G25**

**G13** Stability=3 Size=11 **down pick day10** sub-pattern in downward general pattern (it **split from G18** see dendogram)

**G14** **Stability=9** Size=15     **down pick day 10**     split from **G28**

**G15** **Stability=10** Size=40     **upward pattern;**     **split from G22**

**G16** Stability=3 Size=340     **upward pattern;** **major size node**     **split from G22**

**G17** Stability=7 Size=12     **down pick day 10**     split from **G28**

**G18** **Stability=11** Size=470 down pattern; **major size node** from which merge:G2,G3,G5,G9; G4( dendogram:**G18** )     collagen cluster 67 members are collagen related.

**G19** Stability=3 Size=11      down pick day 10     split from **G28**

**G20** Stability=6 Size=12     high- pick day 6,down-going main pattern, down-pick day12

**G21** Stability=5 Size=27    down pick day 10 ; from G21, splits G14,G17,G19. G21,splits from **G28**

**G22** Stability=3 Size=470    upward pattern& low pick day 10;high pick day6 major size node from which merge:      G6, G7,G10 , G11, G15,G16

**G23** Stability=4 Size=27    upward pick day10;     split's from G25

**G24** Stability=3 Size=14    high day 6, decreasing pattern rest of life

**G25** Stability=3 Size=100    upward pick day10;     split's from G25

**G26** Stability=4 Size=11    oscillatory down pattern, senescence pattern-increase day 12-day15

**G27** Stability=3 Size=12 <mark>up-pick pattern day 10</mark>

**G28** Stability=3 Size=789 <mark>upward pattern</mark> <mark>major size node</mark>

**G29** Stability=5 Size=11 down going pattern - <mark>senescence as pattern;</mark> high expression day 12-day15

**G30** Stability=5 Size=10 down pick day 9, pseudo- <mark>senescence pattern</mark>

**G31** Stability=3 Size=12 <mark>senescence pattern</mark>

**G32** <mark>Stability=11</mark> Size=12 oscillatory pattern

**G33** Stability=4 Size=1941

**G34** Stability=4 Size=1978

**Table 10: Clustering results: list, pattern description.**

**2.3.2 Specific group of genes analysis-Supervised analysis**

Besides performing an un-supervised analysis I was interested in specific groups of genes. To analyze these genes I used a supervised type of analysis. For a better understanding of gene pattern data was normalized using same method described in Section1.

## Germ line enriched genes- pattern analysis

We analyzed a list of ~ 500 genes known to be germ line enriched (Lund et.al. 2002). We found 472 genes out of the 500 on our arrays. We checked the pattern behavior of these genes in entire data set as well as in various lists filtered based on the 'variance threshold-method 'described for my assembly of list of 2000 greatest variance genes.[1] Using such a filtering method I obtained 2 more gene lists of 4500 genes and respectively 1100 genes. We analyzed the gene expression pattern of all 472 germ line enriched gene list find in our data of 18615 genes as well as in the 4500 gene list and in the 1100 gene list. In the list of 4500 genes I identified 220 out of ~ 500 germ line enriched genes and in the list of 1100 genes I've find 51 genes out of the 500 germ line enriched genes. For each list we normalized the gene expression in order to facilitate comparison of the gene patterns. Below is the graph of germ line gene expression pattern corresponding to 472 out of 500 found in 18617 gene list, 220 out of 500 found in 4500 list and 51 gene germ line enriched out of 500 found in 1100 gene list.

---

[1] We computed the variance for each gene per time point. A ranking between all variances has been performed and we choose the first 4500 genes with highest variance. In the same way we identified the list of 1100 genes.

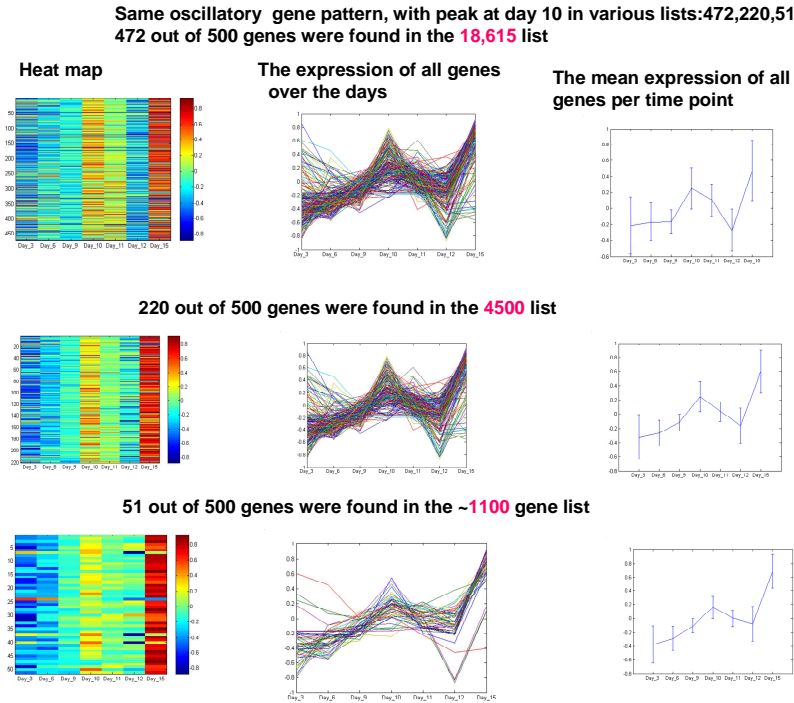

**Fig. 14  Germ line enriched genes.**

**Legend  fig. 14 :**

- First row are the 472 germ line genes found in our raw data in 3 graph representation: heat map, gene expression over time, mean expression of all gene per time point.

- Second raw: 220 germ line genes found in 4500 gene list with highest variation genes in 3 graph representation: heat map, gene expression over time, mean expression of all gene per time point.

- Third raw:  51 germ line genes found in 1100 gene list with highest variation genes. in 3 graph representation: heat map, gene expression over time, mean expression of all gene per time point.

A predominant oscillatory pattern with a 'day 10 peak' can be clearly depicted in all three lists we analyzed. The same pattern we depicted when we clustered the data and find cluster G25. The main biological theme of this cluster was also of germ line genes. Since the genes find in the cluster G25 have a common pattern regulation with germ line genes we might consider that all genes in G25 cluster might have a common loop regulation with germ line enriched genes (see also the comments for cluster G25). Given that this day 10 up-regulated peak was detected when we analyzed directly genes known to be involved with germ line and found also in one of the clusters with a predominantly germ line genes as a biological theme, we might conclude that such pattern is the signature of germ line related genes and of that genes which might have a common loop regulation with germ line enriched genes. Further we might consider that germ line genes have a major role in the transition identified at day 10.

**Tissue specific gene expression analysis**
**Insulin related genes**

**The gene expression pattern of insulin genes suggests involvement in all major processes.**

Given the importance of insulin pathway in aging biology I wanted to get an understanding of the expression of insulin related genes I have on the microarray chips. I identified 23 insulin related genes, see bellow Table 11:

| AFFY_ID | names | rdesc | | | | | | | |
|---------|-------|-------|---|---|---|---|---|---|---|
| ZK75-2_at | ins-2 | q09627 caenorhabditis elegans. probable insulin-like peptide beta-type 2 precursor. 7/1999 | | | | | | | |
| ZK75-3_at | ins-3 | q09628 caenorhabditis elegans. probable insulin-like peptide beta-type 3 precursor. 7/1999 | | | | | | | |
| ZK75- | ins-4 | q09626 caenorhabditis elegans. probable insulin-like peptide beta-type 1 | | | | | | | |

| 1_at | | precursor. 7/1999 |
|---|---|---|
| ZK84-3_at | ins-5 | best hit: p56173 caenorhabditis elegans. putative insulin-like peptide beta-type 6. 7/1998 1.0e-45 89% |
| ZK84-6_at | ins-6 | p56174 caenorhabditis elegans. probable insulin-like peptide beta-type 5 precursor. 7/1998 |
| ZK1251-2_at | ins-7 | q23430 caenorhabditis elegans. probable insulin-like peptide beta-type 4 precursor. 7/1999 |
| C17C3-4_at | ins-11 | q18060 caenorhabditis elegans. probable insulin-like peptide gamma-type 1 precursor. 7/1998 |
| F56F3-6_at | ins-17 | q20896 f56f3.6 protein. 5/2000 |
| T28B8-2_at | ins-18 | ce16518 insulin-like growth factor i like (cambridge) tr:o18149 protein_id:cab03444.1. 0/0 |
| M04D8-1_at | ins-21 | q21507 caenorhabditis elegans. probable insulin-like peptide alpha-type 1 precursor. 7/1998 |
| M04D8-2_at | ins-22 | q21508 caenorhabditis elegans. probable insulin-like peptide alpha-type 2 precursor. 7/1998 |
| M04D8-3_at | ins-23 | q21506 caenorhabditis elegans. probable insulin-like peptide alpha-type 3 precursor. 7/1998 |
| ZC334-3_at | ins-24 | q9u1p6 zc334.3 protein. 5/2000 |
| ZC334-1_at | ins-26 | q9xui9 zc334.1 protein. 11/1999 |
| ZC334-2_at | ins-30 | q9xui8 zc334.2 protein. 11/1999 |
| T10D4-4_at | ins-31 | q9tzf3 t10d4.4 protein. 5/2000 |

| Y8A9A-6_at | ins-32 | q9tyk2 y8a9a.6 protein. 5/2000 |
|---|---|---|
| W09C5-4_at | ins-33 | q9u333 w09c5.4 protein. 5/2000 |
| F52B11-6_at | ins-34 | ce18726 locus:ins-34 (cambridge) protein_id:cab05196.1. 0/0 |
| K02E2-4_at | ins-35 | ce18839 locus:ins-35 (cambridge) protein_id:cab04546.1. 0/0 |
| F08G2-6_at | ins-37 | ce19778 locus:ins-37 (cambridge) protein_id:cab04062.1. 0/0 |
| T13C5-1_at | daf-9 | ce04942 cytochrome p450 (st.louis) tr:q27523 protein_id:aaa80380.1. 0/0 |
| R13H8-1_at | daf-16 | best hit: o16850 fork head-related transcription factor daf-16b. 5/2000 0.0e+00 96% |

**TABLE 11: 23 insulin related genes that are on our array.**

Below is the hit map for the 23 insulin genes-. Besides genes where represented also in a graph with expression of each gene over time in fig.15

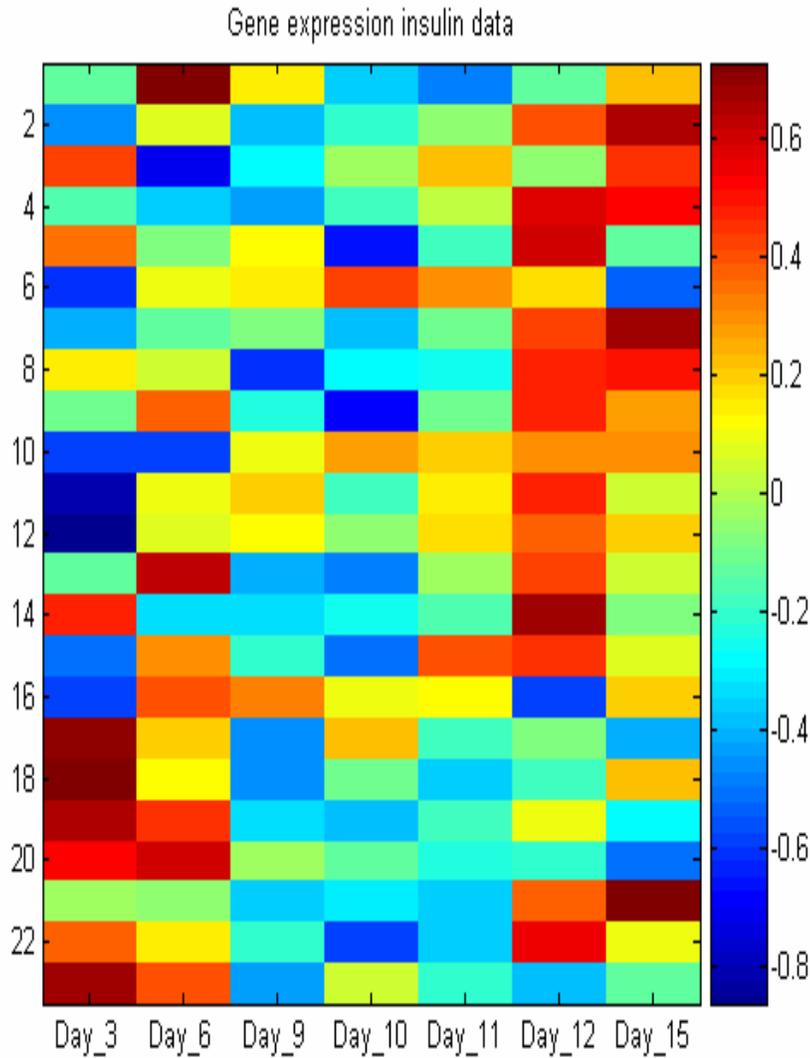

**Legend:** each row is a gene expression, each column is a time point sample.

Red is high level gene expression, blue is low level gene expression. Gene expressions are normalized.

**Fig.15 heat map of the 23 insulin genes**

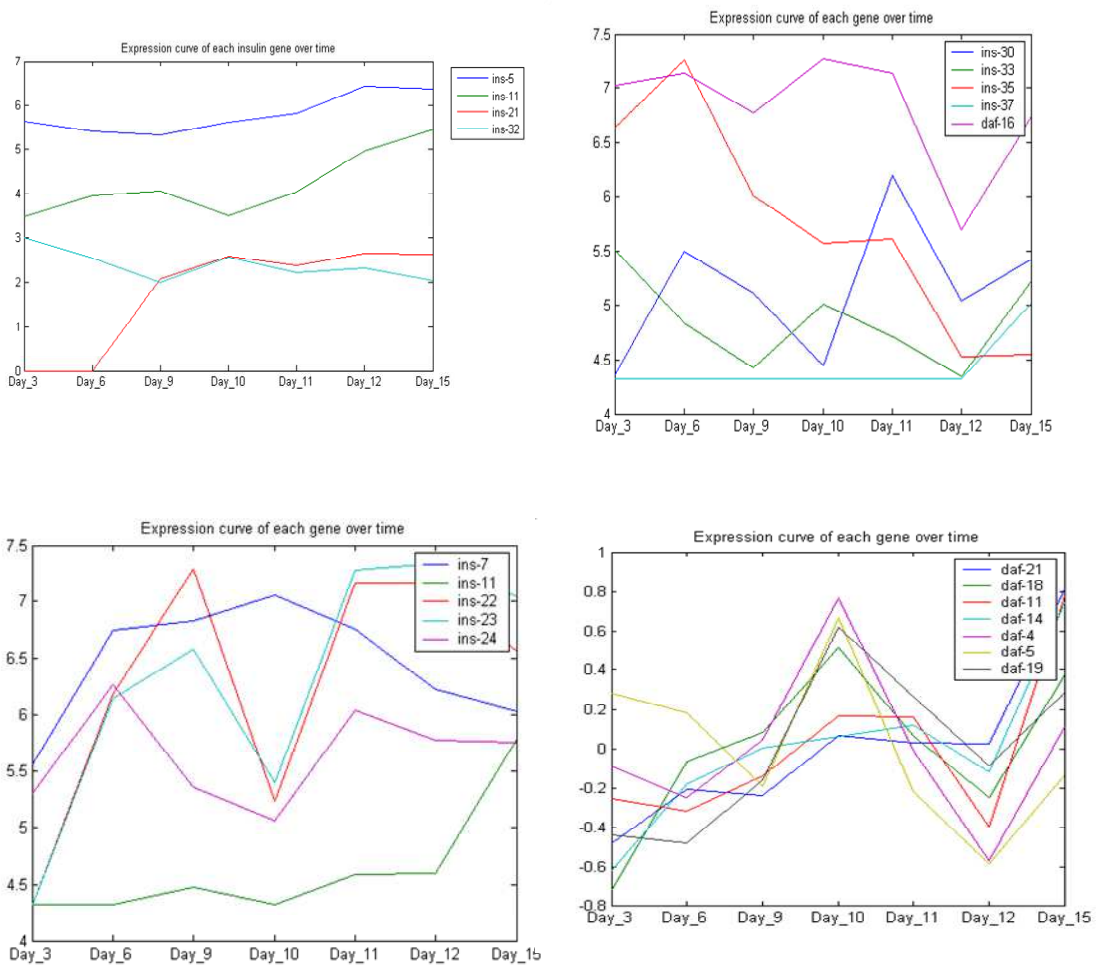

**Fig. 16 expression of insulin related gene out of 23 genes. The 4th fig right down are dauer related gene expressions normalized.**

The first 3 plots from fig 16 are expression profile of insulin related genes, approximate 5 genes per plot. The 4th plot contains dauer related genes found on our array.

Various patterns can be identified in this plot. For example 'day 10 up-peak pattern' found in the cluster G25 can be seen in the dauer related genes at the bottom of fig 10, right side, the genes: *daf-21 daf-18;11;14;4;5;19* and in expression of *ins-7* gene. A down-peak at day 10 can be seen for *ins-22, ins-23, ins-24.*

The insulin genes *ins-5;11;21;32* show an up-regulated pattern.

*ins-35* and *daf-16* show an oscillatory down-regulated pattern and the *ins-30* and *ins-33* show an oscillatory pattern. The *ins-37* has an interesting 'senescence pattern' of steady state over all time points and than an abrupt increase in the expression between day12-15.

Basically, we can identify all 4 patterns found when the general data were clustered for insulin and dauer related genes . This finding can be interpreted as following: if we consider the 4 patterns found in the clustered data as patterns that are describing the main biological processes in this organism, than by observing that insulin genes are expressed in all 4 patterns suggests that insulin genes are involved in all main processes which this nematode undergoes: aging, development, homeostasis. I may further infer that insulin and dauer genes might affect the longevity of *C. elegans* just in an indirect way. The fact that when mutations occur on insulin pathway this in turn affects longevity might be just a signature that insulin pathway actually affects some other vital biological processes which in turn will have an affect on the length of the life time of the *C. elegans*.

- **_daf-21, daf-18, daf-11, daf-14, daf-4, daf-5, daf-19_ and _ins-7_  might share same regulatory loop as**

*daf-2, daf-16* and *ins-7*

Another interesting aspect is that all daf genes mentioned together with *ins-7* have the same oscillatory pattern with a day 10 up-peak. We know from Kenyon results (Nature 424,2003) that when DAF-2 is active, DAF-16 activity is inhibited and *ins-7* is expressed, allowing further DAF-2 activation. When DAF-2 activity is reduced, DAF-16 is activated and *ins-7* expression is inhibited. Our finding that other daf genes: *daf-21, daf-18, 11,14,4,5,19-* share the same pattern with *ins-7* suggests other insulins may participate in the same or similar regulatory loops.

From this survey of 23 insulin related genes, 6 of genes were in our list of 2000 genes, expressed with highest variance. Below is the plot for expression of this genes-see Fig. 17a. The second plot represents the normalized data-see Fig. 17b

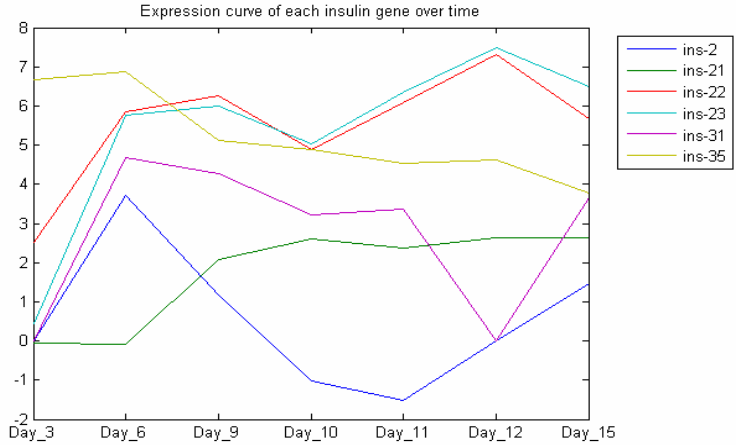

**x-axis: time points; y-axis: gene expressions-not normalized, log2 applied.**

**Fig. 17a-6 insulin genes in our 2000 gene list with highest variance**

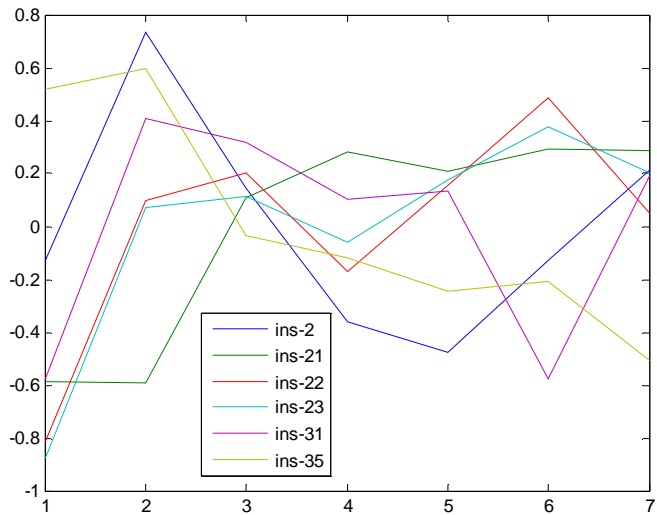

**x-axis: time points; y-axis: gene expressions- normalized, log2 applied.**

**Fig. 17b-6 insulin genes in our 2000 gene list with highest variance-normalized data.**

**Differences in insulin gene patterns might be a direct reflection of the stochastic behavior of gene expression.**

Comparing our data on insulin gene expression with Kenyon results (Nature 424,2003), I note that in the Kenyon data *ins-2* is increased and *ins-21* decreases slightly, whereas in our data, *ins-21* definitely increases and *ins-2* has an oscillatory down-going pattern. Various explanations might be offered for the difference in the patterns, including the differences in the chips and technology used and the difference in the biological strains used. Nevertheless, the pattern difference might also be considered a direct reflection of the stochastic behavior of the gene expressions in *C.elegans*. I will return to these issues of interpreting the comparison between data sets in different labs when I make a careful comparison of our data with other two microarray experiment data.

**Neuronally expressed genes:**

For the neuron expressed genes we focused on a list of approximate 90 genes. The heat map for the expression of these ~ 90 genes, 88 genes to be more precise, can be seen in fig.18. The entire list can be found in Appendix A.

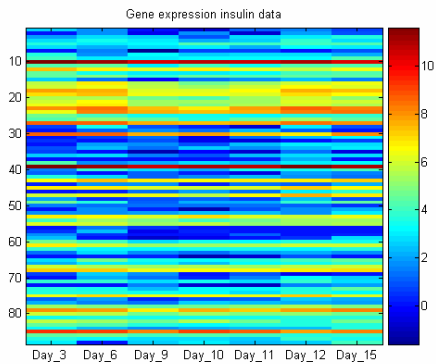

Each row is a gene expression, each column is time point sample.

Red is high level gene expression, blue is low level gene expression. Gene expressions are normalized.

**fig. 18   Heat map of the 88 neurons**

From the graph we can already depict two gene categories: a large number of genes with a relative steady expression value and another category of genes where each of the gene has a distinct pattern of expression.

- **Neuronal related genes might be regulated by environmental cues**

We were interested to find out what genes from the list of ∼ 90 genes expressed in neurons are in our list of 2000 genes that exhibits highest variation.

Out of the ∼ 90 genes I examined, I found 10 genes in the 2000 filtered list based on highest variation. See neurons_table. The gene expression pattern of each neuron related gene appears to be a separate pattern.

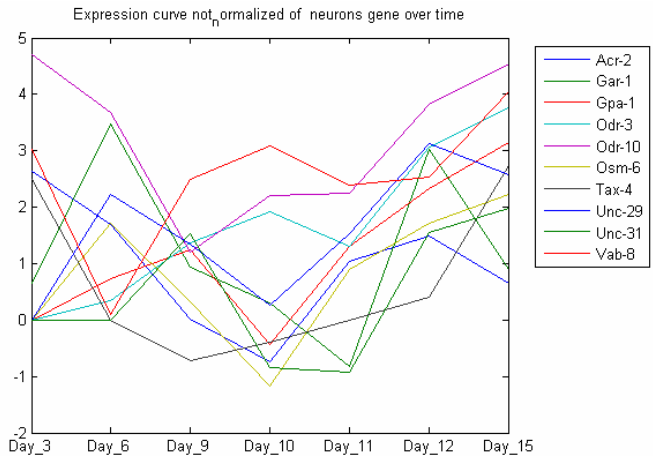

fig. 19   10 neuronal related genes (log2)   out of 88 in the 2000 list; **x-axis: time points; y-axis: gene expressions- normalized, log2 applied.**

This might be a signature for the fact that some neuronal related genes might be regulated by various stimulus and environmental cues over entire life of this organism and that these cues are perceived and integrated in a complex and sophisticated fashion by specific neurons. These 10 genes might be a signature of the neuronal genes that do change over time and consequently impact various processes at various time points.

**'Steady state' expression pattern-a signature consistent with no morphological changes at neuron level.**

We wanted to see if neuronal related genes might have any other pattern. In this sense I enlarged the list of genes with highest variation at 4000. We found 18 more genes out of the 90 genes. They present a distinct clear 'steady' gene expression pattern over the life span of *C. elegans* See fig 20.

# Neurons related genes

## 18 genes out of 88 genes in the list with highest variation

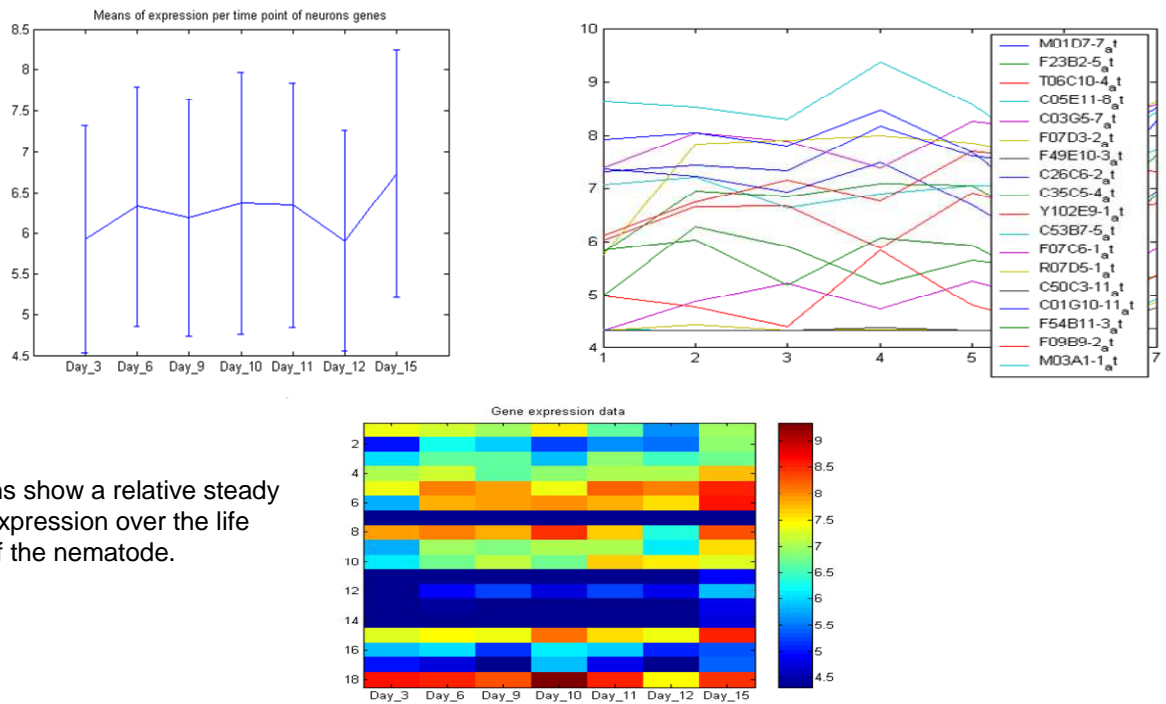

Neurons show a relative steady gene expression over the life span of the nematode.

Fig. upper right: gene expression over time **x-axis: time points; y-axis: gene expressions- normalized, log2 applied.**

**Fig down:** Each row is a gene expression, each column is time point sample.

Red is high level gene expression, blue is low level gene expression. Gene expressions are normalized.

**Fig. 20   neuronal 18 genes out of ~90 genes**

- **Two distinct categories of neuron related genes:**

We concluded that neuronal related genes can be grouped into two distinct categories: one which shows steady expression over the life span of the nematode and another with variable gene expression over time. The steady state pattern of gene expression might be a signature of Driscoll lab hypothesis that neurons don't show significant morphological changes during the life of the nematode. The unchanged group of genes is represented by 18 neuronal expression genes.

On the other hand, the group of 10 genes implies that gene expression in aging animal might be regulated by environmental cues and that these cues are perceived and integrated in a complex and sophisticated fashion by specific neurons. These 10 genes might be a signature of the neuronal genes that do change over time and consequently impact various processes at various time points.

**Genes predicted to impact longevity**

Next we've checked a list of genes considered to be involved in life extension of *C. elegans*. The list of 260 genes was taken from Murphy et. al. 2003, (Nature 424). Approximately one in four genes from the list of life extended genes of 260 genes is included in our list of 2000 genes with highest variation. The heat map for the 58 genes in this graph can be seen in fig 21.

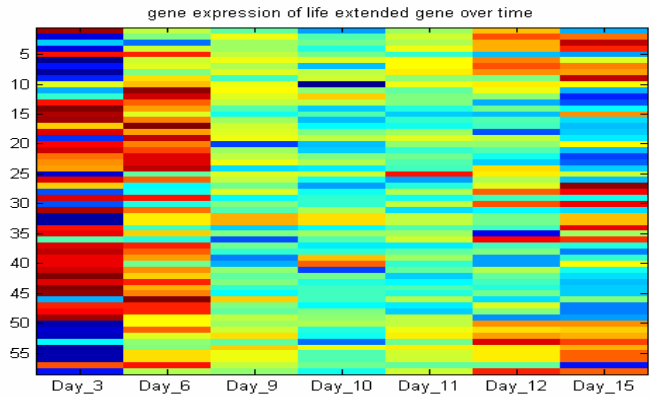

Each row is a gene expression, each column is time point sample.

Red is high level gene expression, blue is low level gene expression. Gene expressions are normalized.

**Fig 21.** Heat map of **58** genes out of 260 **life extended genes** from Murphy et.al. 2003, (Nature 424)

For a complete description of the 58 life-extended genes see Appendix B

- **Life extended genes- regulate key developmental switch, metabolic rate and core processes in general, in accord with evolutionary theory of aging.-two class genes:**

Some of these genes regulate a key developmental switch, while the others control core processes, such as the overall rate of metabolism. These are exactly the kinds of processes predicted to be important to longevity by the evolutionary theory of aging. This theory suggests that competition for metabolic resources between processes such as growth, reproduction and cellular maintenance lies at the heart of the ageing process.

Based on the clustering analysis these genes can be broadly classified into two classes: one of increasing and the other of decreasing in expression pattern.

- **General gene expression pattern suggests that the events are taking place at the beginning of the adulthood day 3-day6 might influence the final days of this organism**.

In our data the 58 life extended genes show a high expression pattern at day 3, day 6, a decrease in expression up to day 12 and than an increase again between day12-15

If we consider the possibility that an increase in expression pattern reflects an increase in the gene activity, than the genes responsible for prolonging the life span of the nematodes have an increase in activities at the beginning of the adulthood and than again at the end of the life of this organism suggesting that the events might take place at the beginning of the adulthood day 3-day6 might determine the final days of this organism.

## Autophagy related genes and their implications in aging studies.

A recent paper provides evidence that macroautophagy is an essential downstream pathway for one of the mutations known to extend life span (A. Melendez, B. Levine, *Science* (2003)) Autophagy, or the degradation of intracellular components by the lysosomal system, was thought for a long time to be a catabolic process responsible for cellular cleanup. However, in recent years, we have learned that autophagy comes in different sizes and shapes, macroautophagy being one of them, and that this cellular maid plays many more roles than previously anticipated. Activation of autophagy is essential in physiological processes as diverse as morphogenesis, cellular differentiation, tissue remodeling, and cellular defense, among others. Furthermore, macrophautophagy participation in different pathological conditions, including cancer and neurodegeneration, is presently a subject of intense investigation. A role in aging has now been added to this growing list of autophagy functions. The activity of different forms of autophagy decreases with age, and this reduced function has been blamed for the accumulation of damaged proteins in old organisms. Research shows that there is much more than trash to worry about when autophagy is not functioning properly. We wanted to investigate the expression gene pattern of authophagy genes in our data set. We identified 7 authophagy related genes that are significant regulated during adult life. Their expression pattern is presented in fig.22:

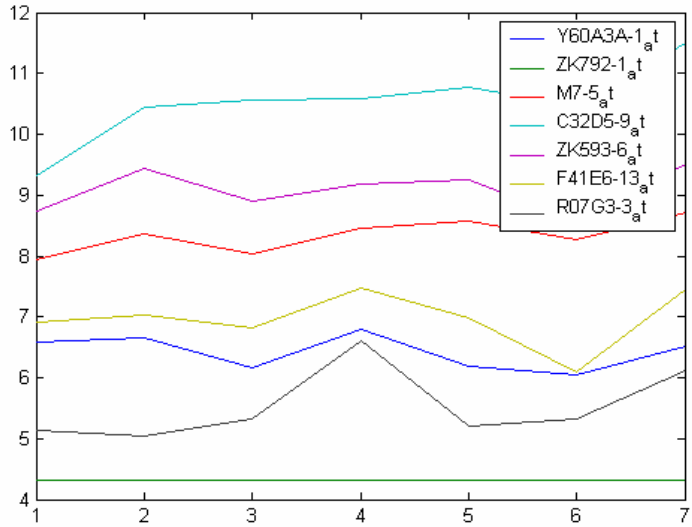

**fig.22 authophagy genes:** gene expression over time; **x-axis: time points; y-axis: gene expressions.**

- **Gene expression pattern suggests a common regulatory loop between certain daf genes and autophagy genes.**

As can be seen in the table 12 bellow and fig. 22 the autophagy related genes have similar pattern and have a role in dauer larval development as well.

| R07G3.3 | npp-21 | authophagy, dauer larval development |
| F41E6.13 | atg-18 | authophagy, dauer larval development |
| ZK593.6 | lgg-2 | not required for dauer larval development or extended life |
| C32D5.9 | lgg-1 | authophagy, dauer larval development |
| M7.5 | atg-7 | authophagy, dauer larval development |
| ZK792.1 | | nematode larval development, autophagy |
| Y60A3A.1 | unc-51 | authophagy, dauer larval development |

Table 12-autophagy related genes

Both dauer formation (a stage of developmental arrest) and adult life-span in *Caenorhabditis elegans* are negatively regulated by insulin-like signaling, but little is known about cellular pathways that mediate these processes. Dauer formation is associated with increased autophagy (A. Melendez, B. Levine et al., Essential role of autophagy genes in dauer development and lifespan extension in *C. elegans. Science* (2003)).

Interestingly, indeed in our data we find that same oscillatory pattern with 'day 10' peak is shared by both dauer genes (see fig 23) and *npp-21*, *atg-18* and *unc-51* autophagy genes.

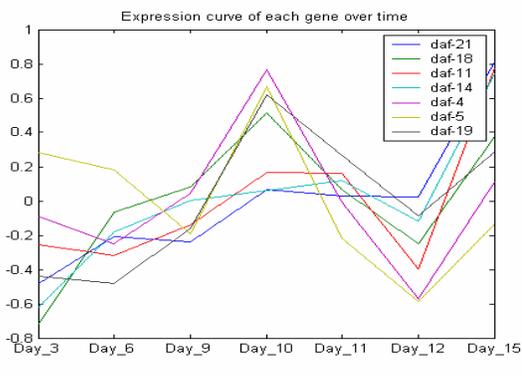

**Fig.23 dauer genes** gene expression over time **x-axis: time points; y-axis: gene expressions- normalized, log2 applied.**

This finding might suggests that *daf-21,18,11,14,4,5,19* and *npp-21*, *atg-18*, *unc-51* may participate in the same or similar regulatory loops.

**<u>Helicases related genes and aging:</u>**

Deficiency in a helicase of the RecQ family is found in at least three human genetic disorders associated with cancer predisposition and/or premature aging. The RecQ helicases encoded by the *BLM*, *WRN* and *RECQ4* genes are defective in Bloom's, Werner's and Rothmund–Thomson syndromes, respectively. Cells derived from individuals with these disorders in each case show inherent genomic instability.

We identified in our data four genes known to be helicase related:

ceWrn him-6 ceRecQ5 ceRecQ4. see fig 24 left. The data presented in fig 24 is normalized for comparison purpose.

Given the importance of the helicase genes in aging related diseases we wanted to see how similar is their expression in one more data set. Lund et. al. performed an experiment similar to our experiment. In spite of some major differences as the type of chip used, time points and strains used, differences which will be stressed later, fundamentally, the design of the experiment has some similarities in the sense that the time points in both data sets cover the entire life span of *C. elegans* and both experiments have replicates, therefore we've checked the helicase gene expression in Lund et. al. data as well, see fig. 24, right.

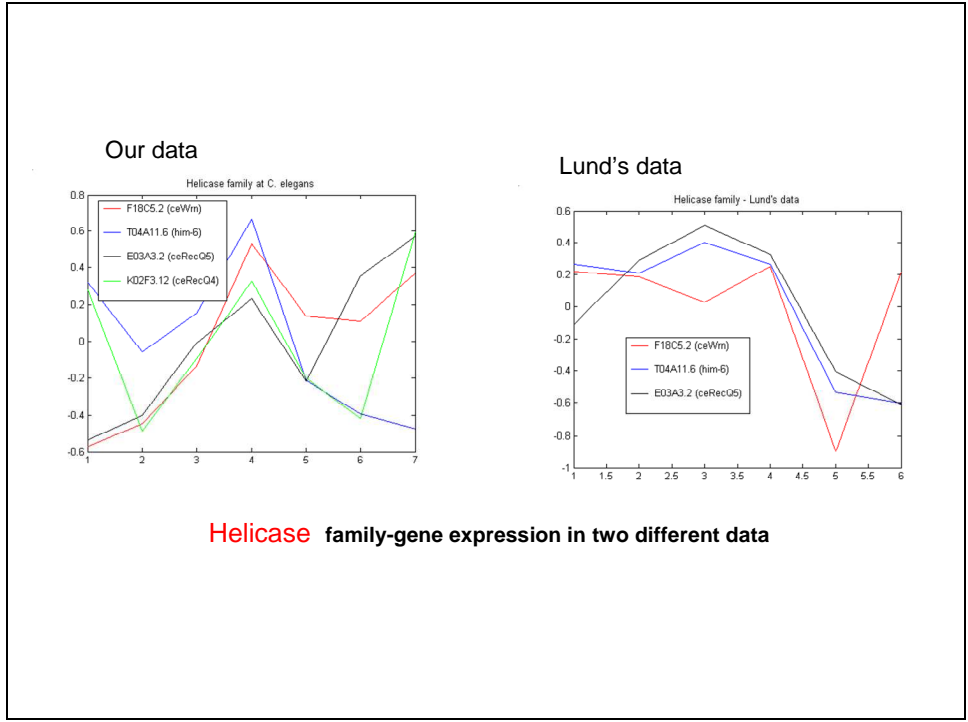

**left: our data;** ox-axis: 7 time points: day3,day6,day9,day10,day11,day11,day15.
oy-axis: normalized gene expression. Log 2 was applied.
**right: Lunda data;** ox-axis: 6 time points: 1/day3 ; 2/day4 ;3/day6-7 ; 4/day9-11 ;
5/day12-14; 6/day16-19; oy-axis: normalized gene expression. Log 2 was applied.
**fig.24 Helicases genes**

The same oscillatory pattern of day 10 can be notice for *him-6* in both data sets. The
*RecQ-5* gene as well as the *ceWrn* gene related with Werner syndrome in humans
show a different pattern in each data set. This might be interpreted as a lack of
robustness of this gene for various stochastic factors that influences *ceWrn* and
*RecQ-5* gene expression. The fact that the gene expression pattern is not repetitive
in the two experiments could be considered a feature of the genes involved with
genomic instability.

DNA helicases are molecular motors that catalyse the unwinding of energetically
unstable structures into single strands and have therefore an essential role in nearly
all metabolism transactions. Defects in helicase function can result in human syndromes in which predisposition to cancer and genomic instability are common features. RecQ helicases are a family of conserved enzymes required for maintaining the genome integrity that function as suppressors of inappropriate recombination. Mutations in RecQ4, BLM and WRN give rise to various disorders characterized by genomic instability and increased cancer susceptibility. One of the signatures of such genes involved in genomic stability might be exactly this inconsistency in gene expression between experiments due to the stochastic factors modulating the expression level of such genes.

## **Muscle related genes and aging**

The behavioral study of ageing nematodes showed a significant decrease in mobility. Age-associated locomotory defects increase progressively in severity over time. Progressive locomotory impairment during *C. elegans* ageing could be the consequence of a decline in muscle function.

For muscle related genes we had several lists we wanted to check in our data. First we checked a list of 60 muscle genes , obtained from Lund et al., 2001, Curr Biol. 12(18):1566-73.

In entire raw data of 18 668  the expression of this group of 60 genes after normalization is as follows: high expression for day 3 to day 6, a relatively steady expression between day 9 to day 12 and again an increase in expression between day 12 to day 15. A complete list of these genes as well as a short description of them it can be seen  Appendix C.  A graphic representation of this genes can be seen in fig. 25 see bellow.

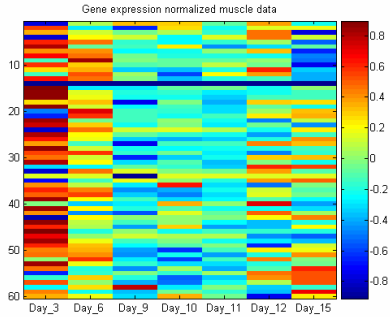

fig. 25  Muscle data_60 genes normalized

We've classified the 60 genes in structural, development/assembly, contraction and anchoring. See  Appendix C .

When we've checked this 60 genes in our list of 2000 genes with highest variation we found 14 genes. See below  their expression pattern in  fig 26.

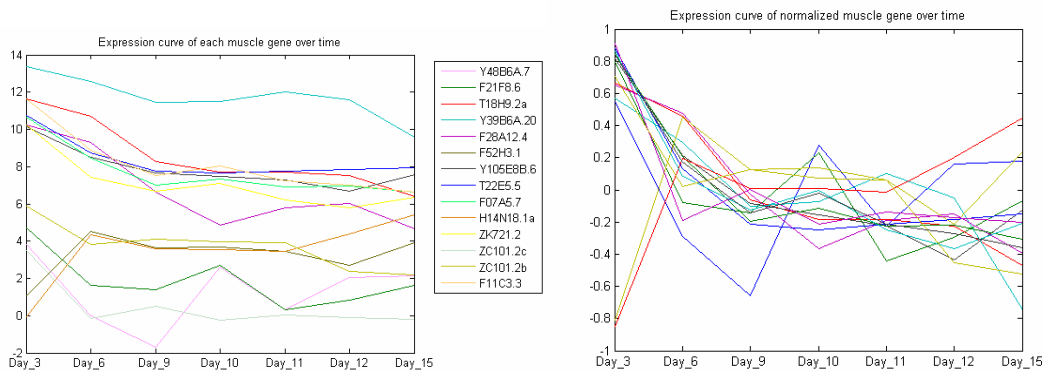

Fig26. a) 23 muscle out of 60 identified in the list with 2000 genes b) same 14 genes normalized.

A relatively steady expression patterns between day 9 to day 12 can be seen in most of the expression patterns of the 14 genes.

When I examined the list of 1283 muscle related  genes proposed by Kim (Roy,Kim et.al, 2002, Nature 418) we found 1187 genes, on our chips among which 111 were within the list of 2000 filtered genes. (see Appendix C)

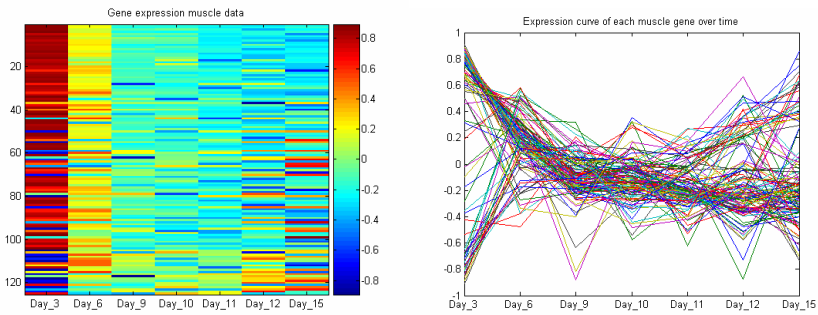

fig. 27    125 genes normalized in common between 2000 list & 1187

As for the previous muscle lists we analyzed, most of the genes are highly expressed at day 3, than a relative steady expression from day 6 to day 12 and than again the expression pattern of this genes become higher. For complete references on the genes name list and a short description of 1187 muscle related genes, 111 muscle related genes see Appendix C.

In conclusion, all muscle expressed genes checked in our data and involved in structural, development/assembly, contraction and anchoring share a similar pattern: high expression day 3, day 6, steady expression at the middle life span of *C. elegans* (day9-12), and again high expression toward the end of life of the organism (day12-15). It is interesting that this pattern is  similar to that of  the long life gene pattern (from Kenyon et. al) previously analyzed. The apparent changes, beginning in mid-life, muscle structure (see Driscoll et.al, Nature 2003) might have as signature at transcriptional level the low expression level between day6-day12. Besides, as later research I've done will show, another clearly pattern muscle related is of decrease in expression level starting with day 3, all the way up to day 15.

We performed a much thorough analysis later on the data muscle gene when we wanted to understand sarcopenia process. Later results will conclude that are actually two distinctive patterns at the muscle level: one of low expression between day 6-12, and another of decrease in expression for all time points between day3-day15. One use of this finding might be if the muscle related group of genes which start to show an increase of the expression level between day 12-15 might be considered as genes with the potential of reverting the sarcopenia.

## Conclusions:

Microarray experiments seek to obtain readouts of gene expression levels over the whole transcriptome . This information can be useful for determining how the transcriptional regulation of genes might coincide, thereby implicating proteins as parts of networks acting together towards a common biological function. Such experiments are particularly useful for complex biological traits that result from the presumed functioning of several molecular pathways. Aging is one such biological trait that apparently incorporates numerous molecular mechanisms underlying environmental stimulus sensing, metabolic regulation, stress responses, reproductive signaling, and transcriptional regulation. Current models of aging emphasize different mechanisms as driving forces behind aging and lifespan determination. However, an integrated understanding of exactly how these mechanisms drive aging has not yet been formulated.

The methods I used for gaining a better understanding of the mechanisms which might underline the aging process where supervised and unsupervised. When interpreting the data, using a supervised approach, I tried to follow the major biological theories currently known that describe aging. In this sense to address the oxidative damage theory of aging, for instance, I've identified stress response genes that exhibit statistically significant changes, and then ask whether the expression patterns of these genes share a common pattern. Also I've looked into insulin and dauer pathway, all being considered important leads in aging studies. Insulins, aging-related gene, dauer-related genes, , autophagy related genes, muscle, neuronal and germline genes all are singled out and their expression profiles examined.

I've investigated and addressed two major aging hypothesis, both being developed in Driscoll's lab; one is pointing out to a major 'crisis' which is going on in the midlife period of time of the organism especially at the muscle level, which might be critical for determining the ultimate lifespan of that animal, the other is underlining the idea that aging must be understood as an stochastic process due to stochastic cues acting on the organism over the entire lifespan of that organism. The implications of aging as a stochastic process can be seen at the transcription level in various specific group of genes as I've pointed out when I've analyzed insulin or neuronal related genes.

**Aknowledgements:**

This work would have not been possible without the constant discussions with Prof. Monica Driscoll. The microarray raw data are of Driscoll lab. Also I have to acknowledge Prof. Eytan Domany who helped me understand his clustering method during a stage at Weizmann Institute, Institute of Complex Systems. Also, his student at the time, Uri Einav. Nevertheless I have to acknowledge Beate Hartmann,, Christophe Grundschober, and Patrick Nef from Roche who helped providing the raw data to Driscoll Lab. Most of this work was supported by a grant of Prof Monica Driscoll.